\documentclass[12pt]{article}
\usepackage{amsmath,amssymb,epsfig}
\makeatletter \@addtoreset{equation}{section} \makeatother
\renewcommand{\theequation}{\thesection.\arabic{equation}}

\newcommand{\ba}{\begin{array}}
\newcommand{\ea}{\end{array}}
\newcommand{\beq}{\begin{equation}}
\newcommand{\eeq}{\end{equation}}
\newcommand{\bea}{\begin{eqnarray}}
\newcommand{\eea}{\end{eqnarray}}

\def\bce{\begin{center}}
\def\ece{\end{center}}

\def\nonu{\nonumber}

\def\be{\beta}

\def\La{\Lambda}

\def\R{{\bf R}}

\def\eps6{{\displaystyle \mathop{\epsilon}^{6}}{}}

\def\nab6{{\displaystyle \mathop{\nabla}^{6}}{}}

\def\ft#1#2{{\textstyle{\frac{\scriptstyle #1}{\scriptstyle #2}}}}
\def\fft#1#2{\frac{#1}{#2}}
\def\0{{\sst{(0)}}}
\def\1{{\sst{(1)}}}
\def\2{{\sst{(2)}}}
\def\3{{\sst{(3)}}}
\def\4{{\sst{(4)}}}
\def\5{{\sst{(5)}}}
\def\6{{\sst{(6)}}}
\def\7{{\sst{(7)}}}
\def\8{{\sst{(8)}}}

\def\td{\tilde}
\def\nnn{\nonumber}
\def\CP{{{\mathbb C}{\mathbb P}}}

\def\ba{\begin{array}}
\def\ea{\end{array}}
\def\beq{\begin{equation}}
\def\eeq{\end{equation}}
\def\be{\begin{equation}}
\def\ee{\end{equation}}

\def\eps{\epsilon}

\def\ba{\begin{array}}
\def\ea{\end{array}}
\def\beq{\begin{equation}}
\def\eeq{\end{equation}}
\def\be{\begin{equation}}
\def\ee{\end{equation}}

\def\eps{\epsilon}

\newcommand{\bean}{\begin{eqnarray*}}
\newcommand{\eean}{\end{eqnarray*}}

\begin{document}
\thispagestyle{empty} \addtocounter{page}{-1}
\begin{flushright}
{\tt hep-th/0508075}
\end{flushright}

\vspace*{1.3cm} \centerline{ \Large \bf Deformations of Flows from
Type IIB Supergravity} \vspace*{1.5cm} \centerline{{\bf Changhyun
Ahn $^{\ast,\dagger}$ and {\bf Justin F.
V\'azquez-Poritz}$^{\ast,\ddagger}$}} \vspace*{1.0cm}
\centerline{\it  $^{\ast}$ School of Natural Sciences, Institute
for Advanced Study,}\vspace{.1pt} \centerline{\it Einstein Drive,
Princeton NJ 08540, USA}

\vspace{10pt}

\centerline{\it $^{\dagger}$ Department of Physics, Kyungpook
National University,}\vspace{.1pt} \centerline{\it Taegu 702-701,
Korea}

\vspace{10pt}

\centerline{\it $^{\ddagger}$ Department of Physics, University of
Cincinnati,}\vspace{.1pt} \centerline{\it Cincinnati OH 45221-001,
USA } \vspace*{0.8cm} \centerline{\tt ahn@knu.ac.kr, \qquad
jporitz@physics.uc.edu} \vskip2cm

\centerline{\bf Abstract} \vspace*{0.5cm}

We consider supersymmetric $SL(3,R)$ deformations of various type
IIB supergravity backgrounds which exhibit flows away from an
asymptotically locally $AdS_5\times S^5$ fixed point. This
includes the gravity dual of the Coulomb branch of ${\cal N}=1$
super Yang Mills theory, for which the deformed superpotential is
known. We also consider the gravity duals of field theories which
live on various curved backgrounds, such as Minkowski$_2\times
H^2$, $AdS_3\times S^1$ and $\R\times S^3$. Some of the deformed
theories flow from a four-dimensional ${\cal N}=1$ superconformal
UV fixed point to a two-dimensional $(2,2)$ superconformal IR
fixed point. We study nonsupersymmetric generalizations of the
deformations of the above Coulomb branch flows.

\baselineskip=18pt
\newpage
\renewcommand{\theequation}
{\arabic{section}\mbox{.}\arabic{equation}}

\section{Introduction}

Compactification of type IIB supergravity on a two-torus has an
$SL(3,R)$ symmetry which can be used for generating new solutions.
This has recently been applied to finding the type IIB
supergravity background which, via the AdS/CFT correspondence
\cite{Malda1997,GKP,Witten1998}, corresponds to marginal
deformations of ${\cal N}=4$ super Yang-Mills theory \cite{LM}.
The undeformed theory has an isometry group which includes $U(1)
\times U(1)$. The deformation on the gravity side can be matched
to an exactly marginal operator in the field theory, providing a
holographic test of the methods of Leigh and Strassler \cite{LS}.

The solution-generating technique can be outlined as follows.
First, T-dualize along one of the $U(1)$ directions to type IIA
theory. Lifting the solution to eleven dimensions provides a third
direction which is associated with a $U(1)$ symmetry. One can now
apply an $SL(3,R)$ rotation along these $U(1)^3$ directions.
Dimensionally reducing and T-dualizing along shifted directions
yields a new type IIB solution. This procedure can be applied to
any solution that has an isometry group which contains $U(1)\times
U(1)$. If in addition to this symmetry there is a $U(1)$
R-symmetry, then the deformed solution preserves ${\cal N}=1$
supersymmetry.

This method has also been applied to the marginal deformations of
the conifold superconformal field theory \cite{LM}, as well as the
superconformal theories associated to the $Y^{p,q}$ \cite{LM} and
$L^{p,q,r}$ \cite{AV} manifolds\footnote{The Sasaki-Einstein
spaces $Y^{p,q}$ and $L^{p,q,r}$ were found in \cite{Ypq1,Ypq2}
and \cite{Lpqr1,Lpqr2}, respectively.}. Furthermore, the gravity
duals of non-conformal theories, such as those which exhibit
renormalization group flows, can also be deformed in this manner.
In particular, deformations of the Klebanov-Strassler solution
\cite{klebstrass} and the 5-brane wrapped on a 2-sphere
\cite{maldnun} have been considered in \cite{LM} and \cite{GN},
respectively\footnote{For the latter case, deformations which
incorporate one internal direction and one worldvolume direction
were also considered. These correspond to dipole deformations
\cite{GN}.}. These solutions are useful for understanding the
gravity dual of confining gauge theories.

Marginal deformations of eleven-dimensional supergravity solutions
with $U(1)^3$ symmetry have also been studied. The
solution-generating technique is analogous to the above type IIB
cases. One begins with a solution that has $U(1)^3$ non-R
symmetry. Then one reduces to ten dimensions and T-dualizes to
type IIB theory, where one can perform an $SL(2,R)$ transformation
with the remaining two $U(1)$ directions and then dualize back to
eleven dimensions along the shifted directions\footnote{In fact,
there is also a direction associated with the $U(1)$ R-symmetry.
This can be used to perform a multiple chain of dualities and
$SL(3,R)$ transformations resulting in four deformation
parameters. However, the deformed solution would be
nonsupersymmetric. One could consider an eleven-dimensional
solution with a large enough isometry group such that
multiple-parameter deformations preserve supersymmetry.}. This has
been done for initial solutions with geometries of the form
$AdS_4\times X^7$, which correspond to three-dimensional
superconformal field theories. The case in which $X^7$ is a
7-sphere was considered in \cite{LM}, for which the deformed gauge
theory has ${\cal N}=2$ supersymmetry and the exactly marginal
deformation operator was identified. The deformed field theory
also has ${\cal N}=2$ supersymmetry if $X^7$ is one of a countably
infinite number of Sasaki-Einstein spaces\footnote{Countably
infinite classes of seven-dimensional Sasaki-Einstein spaces have
been found in \cite{GMSW03,CLPP,Lpqr1}.} \cite{AV} and has ${\cal
N}=1$ supersymmetry if $X^7$ is weak $G_2$ or tri-Sasakian
\cite{GLMW}.

In this paper, we consider deformations of type IIB solutions
which are asymptotically locally $AdS_5\times S^5$. This implies
that the UV limit of the dual gauge theory is the Leigh-Strassler
deformation of ${\cal N}=4$ super Yang-Mills theory. There is a
renormalization group flow away from this conformal fixed point.
We consider type IIB solutions which are related to solutions of
the $U(1)^3$ truncation of five-dimensional gauged supergravity by
using the dimensional reduction ans\"{a}tz given in
\cite{tenauthors}. This implies that, when lifted to
ten-dimensional type IIB theory, the isometries includes a
$U(1)^2$ symmetry with which to deform the solution while
maintaining a $U(1)$ R-symmetry. That is, at every point along the
flow there is locally an $S^5$ which can be written as a constant
$U(1)$ bundle over $\CP^2$. It is this $U(1)$ direction which
corresponds to the $U(1)$ R-symmetry after the deformations.
Non-vanishing $U(1)$ gauge fields in the five-dimensional
solutions correspond to nonzero $R$-symmetry currents in the dual
gauge theory. Although we have not explicitly checked that the
deformed solutions satisfy the BPS equations for type IIB theory,
since the direction corresponding to the $U(1)$ R-symmetry is not
involved in these deformations, we expect that the deformed
solutions should preserve a fraction of supersymmetry everywhere.

This paper is organized as follows. In the next section, we
consider deformations of a certain region along Coulomb branch of
${\cal N}=4$ super Yang Mills theory at zero temperature,
described by D3-branes distributed uniformly on an ellipsoidal
shell \cite{larsen}. Although this Coulomb branch is maximally
supersymmetric, conformal symmetry is broken-- to be regained only
at the UV fixed point. This static gravity solution arises in the
extremal limit of rotating black D3-branes. The three rotational
parameters $\ell_i$ of the rotating branes become distributional
parameters of the ellipsoidal shell. While a stack of coincident
D3-branes corresponds to the superconformal origin of the moduli
space, moving some of these branes away from each other
corresponds to giving expectation values to certain scalar fields.
The marginal deformation of $AdS_5\times S^5$ considered in
\cite{LM} lies in the UV limit of the deformation of this
solution. The introduction of scalar expectation values does not
change the superpotential. After the deformation, the
superpotential along the Coulomb branch flow is still the same as
at the UV fixed point. That is, the deformation yields phases of
chiral superfields in the superpotential.

In section 3, we consider deformations of twisted field theories
\cite{twisted}. These theories are twisted since they live on a
curved space and are coupled to an $SO(6)$ gauge field. They are
described by D3-branes wrapped on a (quotiented) hyperbolic
2-cycle \cite{MN1}. The corresponding geometry is nonsingular and
remains so after the deformations. In the UV limit, the initial
geometry is locally $AdS_5\times S^5$. The dual field theory
exhibits a flow ``across dimensions'' from a four-dimensional
superconformal field theory to a two-dimensional superconformal
field theory. While the initial two-dimensional field theory has
either $(4,4)$ or $(2,2)$ supersymmetry, depending on how the
2-cycle is embedded within the rest of the internal space, the
deformed two-dimensional field theory has $(2,2)$ supersymmetry.

In section 4, we deform various other solutions which are
asymptotically locally $AdS_5\times S^5$. All of these solutions
were initially obtained in five-dimensional gauged supergravity
and then lifted to ten-dimensional type IIB theory. In five
dimensions, these solutions correspond to a supersymmetric $AdS_5$
black hole \cite{reall1,reall2}, a deformation (in the sense of
adding $U(1)$ gauge fields) of $AdS_5$
\cite{gauntlett,gauntlett1}, an $AdS_5$ magnetic string
\cite{sabra,KS} and a flux-brane\footnote{This solution was
recently renamed an anti-bubble in \cite{jones}.} \cite{LPV}. As
we will discuss, some of these solutions are related to the
solutions considered in sections 2 and 3 and have interesting
holographic interpretations.

In section 5, we will discuss deformations which do not preserve
any supersymmetry. In this case, there are two possibilities:
either the initial solution has supersymmetries which are broken
by the deformations or else the initial solution is itself
nonsupersymmetric. As examples of both scenarios, we consider a
couple of generalizations of the ${\cal N}=4$ Coulomb branch flow
\cite{larsen} discussed in section 2. First we discuss an ${\cal
N}=1$ flow \cite{KW,GW} which asymptotes to the ${\cal N}=4$
Coulomb branch flow in the UV limit. This solution also
encompasses the ${\cal N}=1$ Pilch-Warner flow
\cite{PWfeb2000,PW1}. Since the deformation involves the direction
associated with the $U(1)$ R-symmetry, the resulting solution is
nonsupersymmetric. To illustrate the second possibility, we
consider a nonsupersymmetric generalization of the ${\cal N}=4$
Coulomb branch flow, which corresponds to the gauge theory being
at finite temperature \cite{larsen}. In this case, there are
actually three directions which can be used for the deformations.
After pairing these directions up with a chain of dualities and
shifts, this yields a solution with six deformation parameters
which, in the UV limit, reduces to the solution discussed in
\cite{Frolov,Frolov:2005ty,Frolov:2005iq}. There, it was found
that a three-parameter case of the latter solution has an
associated string sigma model which is integrable.
Nonsupersymmetric deformations have also considered in \cite{GN}.
Since these deformed backgrounds were obtained from solutions of
type IIB supergravity by a string of dualities and shifts, we
expect that they are also solutions to the type IIB equations of
motion. However, we have not explicitly checked this.

In section 6, we present some suggestions for further directions.
We have also included some details of the more complicated
solutions in the appendices.

\section{Supersymmetric deformation of the Coulomb branch flow}

Rotating black D3-branes\footnote{Spinning branes were first
constructed in \cite{youm}.} have three independent $SO(6)$
rotational parameters $\ell_i$. In the extremal limit, rotating
black D3-branes reduce to a static and continuous D3-brane
distribution. The three parameters $\ell_i$ now specify the
ellipsoidal shape of the D3-brane distribution \cite{larsen}. In
five dimensions, the near-horizon limit of this solution
corresponds \cite{gubser,tenauthors} to a supersymmetric, charged
singularity which is asymptotically $AdS_5$\footnote{This solution
is usually referred to as a charged $AdS_5$ black hole, which is a
misnomer since there is no event horizon around the singularity.}
\cite{behrndt}. The ellipsoidal shell containing the uniform
distribution of D3-branes tends to be singular, which could
represent a phase transition at the associated scale of the RG
flow \cite{larsen}. Therefore, the supergravity description breaks
down around this region.

To better understand this configuration, one can first consider
the superconformal origin of the moduli space, which corresponds
to a single stack of coincident D3-branes. Since these D3-branes
are each in a BPS state, moving them away from each other does not
require any additional energy. Therefore, any distribution of
these branes is still maximally supersymmetric. However, with the
exception of the special case in which the stack is expanded out
to a perfectly spherically symmetric distribution, the conformal
invariance of the theory is generally broken. Nevertheless, the
geometry asymptotes to $AdS_5\times S^5$ at the UV fixed point,
where conformal symmetry is regained. Moving these branes away
from each other has another interesting effect. That is, at the
superconformal point there are six scalar fields which transform
in the vector representation of the $SO(6)$ R-symmetry and the
adjoint representation of $SU(N)$, where $N$ is the number of
D3-branes. As branes are moved, these scalars take on expectation
values. Therefore, the resulting supergravity solution corresponds
to a certain region along Coulomb branch of ${\cal N}=4$ super
Yang Mills theory at zero temperature \cite{larsen}.

The introduction of scalar expectation values does not change the
superpotential of the theory. After the deformation, the
superpotential along the Coulomb branch flow is the same as the
marginal deformation of $AdS_5\times S^5$ \cite{LM}, which lies at
the UV fixed point. Namely, the deformed superpotential is given
by
\bea W_{\rm{marg}} = \mbox{Tr} \left( \Phi_1 \Phi_2 \Phi_3 -q
\Phi_2 \Phi_1 \Phi_3 \right), \qquad q^n =1\,,
\label{marginalsuperpot} \eea
which in the ${\cal N}=1$ language is a $q$-deformation of the
superpotential preserving ${\cal N}=1$ supersymmetry \cite{LS}
(See also \cite{BLjan,BJL,BJLjune}). The moduli space of vacua
depends on $q$ and the super Yang-Mills coupling is a function of
$q$. For $q=1$, the ${\cal N}=4$ theory can be described by an
${\cal N}=1$ theory with three adjoint chiral superfields $\Phi_i$
coupled through the superpotential $W_{\rm{marg}}(q=1)$.

We will now consider the near-horizon limit of the aforementioned
distribution of D3-branes in detail. The metric and 4-form
potential can be written as\footnote{We have used the solution
given in \cite{KW} in terms of the notations in \cite{larsen},
except that we have taken $\theta\rightarrow \alpha+\pi/2$,
$\phi\rightarrow\theta$ in order to match the deformation of
$AdS_5\times S^5$ in \cite{LM} in the UV limit. Note that the
expression in \cite{KW} corrects a mistake from that of
\cite{larsen}. Also, the $\psi$ rotation generates $U(1)_R$
symmetry as in \cite{LM}. We use the simple notation $c_{\theta}
\equiv \cos\theta$ and $s_{\theta}\equiv \sin\theta$ and so on.}
\cite{larsen}
\bea ds_{10}^2 &=& f^{-1/2} \fft{r^2}{R^2}
(-dt^2+dx_1^2+dx_2^2+dx_3^2) +f^{-1/2} \fft{R^2}{r^2}
\fft{dr^2}{L_1 L_2 L_3}\nnn\\
&+& f^{1/2} R^2 \Big[ k^{-1}\, [d\alpha+\fft{k}{4}
(L_2-L_3)s_{2\alpha} s_{2\theta}\,d\theta]^2+kf^{-1}
s_{\alpha}^2\,d\theta^2\nnn\\ &+& L_1 c_{\alpha}^2
(d\psi-d\varphi_2)^2 + L_2 s_{\alpha}^2 s_{\theta}^2
(d\psi-d\varphi_1)^2 +L_3 s_{\alpha}^2 c_{\theta}^2
(d\psi+d\varphi_1+d\varphi_2)^2\Big] \,,\nnn\\
F_{(5)} &=& dC_{(4)}+\ast\, dC_{(4)}\,,\qquad
C_{(4)} = \Big( f^{-1} \fft{r^4}{R^4} -1\Big) \,dt\wedge
dx_1\wedge dx_2\wedge dx_3\,,
\label{KLT}
\eea
where $R=(4\pi g_s N)^{1/4}$. We have introduced the following
quantities
\be f^{-1} = (L_1^{-1} c_{\alpha}^2+L_2^{-1} s_{\alpha}^2
s_{\theta}^2+L_3^{-1} s_{\alpha}^2 c_{\theta}^2)L_1 L_2 L_3
\,,\quad g^{-1} = L_2 s_{\theta}^2+L_3 c_{\theta}^2 \,,\quad
k^{-1}=L_1 s_{\alpha}^2+g^{-1} c_{\alpha}^2, \label{fgk} \ee
and
\be L_i = 1+\fft{\ell_i^2}{r^2}\,. \label{L} \ee
This geometry approaches $AdS_5\times S^5$ as $r\rightarrow
\infty$, in which case $f, g, k$, and $L_i$ go to 1. The vector
representation ${\bf 6}$ of $SO(6)$ with respect to the full
Cartan subgroup $SO(2)^3$ decomposes as $({\bf 2}, {\bf 1}, {\bf
1})\oplus ({\bf 1}, {\bf 2},{\bf 1})\oplus ({\bf 1},{\bf 1},{\bf
2})$ \cite{BS99}. The $SO(2)^3$ isometry group of above geometry
corresponds to $U(1)^2$ global symmetry generated by $\varphi_1$
and $\varphi_2$ and $U(1)_R$ $R$-symmetry generated by $\psi$ of
the dual field theory.

Let us write the solution in the form given by equation (A.7) in
\cite{LM}. That is, the metric is written such that the $T^2$
symmetry along the $\varphi_1$ and $\varphi_2$ directions is
explicit:
\bea ds_{IIB}^2 &=& F\Big[ \fft{1}{\sqrt{\Delta}}(D\varphi_1-C
D\varphi_2)^2 +\sqrt{\Delta} D\varphi_2^2\Big] +F^{-1/3} g_{\mu
\nu} dx^{\mu} dx^{\nu} \,,\nonu\\ F_{(5)} &=& R^{-4}\,d(f^{-1}
r^4)\,dt\wedge dx_1\wedge dx_2\wedge dx_3+ d{\cal A}_{(2)}\wedge
D\varphi_1\wedge D\varphi_2\,, \label{metric} \eea
where
\bea F^{-1/3} g_{\mu \nu} dx^{\mu} dx^{\nu} &=& f^{-1/2}
\fft{r^2}{R^2} (-dt^2+dx_1^2+dx_2^2+dx_3^2) +f^{-1/2}
\fft{R^2}{r^2}
\fft{dr^2}{L_1 L_2 L_3} \nnn\\
&+& f^{1/2} R^2 \Big[ k^{-1}\,
[d\alpha+\fft{k}{4}(L_2-L_3)s_{2\alpha}
s_{2\theta}\,d\theta]^2\nnn\\
&+ & kf^{-1} s_{\alpha}^2\,d\theta^2 +\fft{9}{4} ghL_1 L_2 L_3
s_{2\alpha}^2 s_{2\theta}^2\,d\psi^2 \Big]\,. \eea
We have explicitly taken the Hodge dual in $F_{(5)}$ to find
\bea d{\cal A}_{(2)} &=& \fft34 R^4 \fft{f^2}{r} s_{2\alpha}
s_{2\theta} \,d\psi\wedge \Big( \fft{L_1L_2L_3}{r^2}
\partial_r (f^{-1} r^4) s_{\alpha}^2\,
d\alpha\wedge d\theta + k f^{-1}
\partial_{\alpha} (f^{-1}) s_{\alpha}^2\,
dr\wedge d\theta\nnn\\ &+& [\fft{L_3-L_2}{4} s_{2\alpha}
s_{2\theta} \partial_{\alpha}(f^{-1})+\fft{1}{k}
\partial_{\theta}(f^{-1})]\, dr\wedge [d\alpha+\fft{k}{4}(L_2-L_3)
s_{2\alpha} s_{2\theta}\,d\theta] \Big)\,. \label{da2} \eea
Note that $d{\cal A}_{(2)}$ reduces to $12 R^4 c_{\alpha}
s_{\alpha}^3 s_{\theta} c_{\theta} d \psi \wedge d \alpha \wedge d
\theta$ in asymptotic limit $r \rightarrow \infty$, which is
consistent with how \cite{LM} expressed the potential $C_{(4)}$ in
the $AdS_5\times S^5$ solution. In the above, we have introduced
the connection one-forms
\bea D\varphi_1 & \equiv & d\varphi_1+A_{\psi}^1 d\psi
\,,\qquad D\varphi_2 \equiv d\varphi_2+A_{\psi}^2 d\psi\,,\nnn\\
A_{\psi}^1 &=&- g L_3 c_{\theta}^2 (2gL_2 L_3 s_{\alpha}^2
s_{2\theta}^2-4L_1 c_{\alpha}^2)h\,+g(L_3 c_{\theta}^2-L_2
s_{\theta}^2)\,,\nnn\\ A_{\psi}^2 &=& (2gL_2 L_3 s_{\alpha}^2
s_{2\theta}^2-4L_1 c_{\alpha}^2)h\,,
\label{covariant}
\eea
where
\bea
 h^{-1} = 4L_1 c_{\alpha}^2+gL_2 L_3 s_{\alpha}^2 s_{2\theta}^2.
\label{h}
\eea
By comparing the metric (\ref{KLT}) with
(\ref{metric}),
the metric functions are given by
\be F(r,\theta,\alpha) = \fft{f^{1/2} R^2 s_{\alpha}}{2\sqrt{g
h}}\,,\qquad \Delta(r,\theta,\alpha)=\fft{g}{4h
s_{\alpha}^2}\,,\qquad C(r,\theta)=-gL_3 c_{\theta}^2\,
\label{metricfunction}
\ee
where the functions $f,g$, and $h$ are given in (\ref{fgk}) and
(\ref{h}).

We will now use the deformation formulae presented in the appendix
of \cite{LM}. The $SL(3,R)$ transformation can be described by the
matrix
\bea \Lambda = \left(\begin{array}{ccc}
1&  \gamma & 0 \\
0 &1& 0\\
0& \sigma & 1
\end{array}\right)\,.
\eea
This transformation is parameterized by the real parameters
$\gamma$ and $\sigma$, where $\beta=\gamma-\tau_s\,\sigma$ is
related to the aforementioned $q$ by $q=e^{2\pi i \beta}$. In
general, $\tau_s$ is a complex structure parameter that is related
to the gauge coupling and theta parameter of the dual gauge
theory.

The transformation of various scalar fields can be obtained by
constructing $3 \times 3$ matrix $g^T$ \cite{LM} where
\bea g^T= \left(\begin{array}{ccc}
e^{-\phi/3}\,F^{-1/3}& 0& 0 \\
0 &e^{-\phi/3}\,F^{2/3}& 0\\
0&0& e^{2\phi/3}\,F^{-1/3}
\end{array}\right)
\left(\begin{array}{ccc}
1&B_{12} &0\\
0&1&0\\
\chi_{\rm axion}  & -C_{12}+\chi_{\rm axion} B_{12} & 1
\end{array}\right).
\label{gt}
\eea
We choose the dilaton and axion fields to be vanishing: $\phi_{\rm
dilaton}=\chi_{\rm axion}=0$. Then $\tau_s$ is simply $i$ and
$\beta$ reduces to $\gamma-i \,\sigma$. However, in general the
dilaton and axion can be specified by nonvanishing constants.

Under the $SL(3,R)$ transformation $M=g g^T$ goes to $M'=\La M
\La^T$ and the corresponding $g^T$ transforms as \cite{LM}
\bea g^{\prime T}= \left(\begin{array}{ccc}
\fft{1}{\sqrt{GH}F^{1/3}}& 0& 0 \\
0 &\sqrt{G}F^{2/3}& 0\\
0&0&\sqrt{H} F^{-1/3}
\end{array}\right)
\left(\begin{array}{ccc}
1& \gamma\,G\,F^2 &0\\
0&1&0\\
\gamma\sigma\,F^2 H^{-1} & \sigma\,F^2 H^{-1} &1
\end{array}\right),
\label{g't}
\eea
where
\bea H \equiv 1+ \sigma^2 F^2, \qquad G^{-1} \equiv 1+ (\gamma^2 +
\sigma^2) F^2\,.
\eea
Also, the ${\cal A}_{(2)}$ term in the initial five-form field
strength induces additional terms in the deformed two-forms $B$
and $C^{(2)}$ as follows \cite{LM}:
\bea \left(\begin{array}{c}
\fft12 {\td c}_{\mu\nu}\,dx^{\mu}\wedge dx^{\nu}  \\
{\cal A}_{(2)}  \\
\fft12 {\td b}_{\mu\nu}\,dx^{\mu}\wedge dx^{\nu}
\end{array}\right) & \longrightarrow &
\Lambda \left(\begin{array}{c}
\fft12 {\td c}_{\mu\nu}\,dx^{\mu}\wedge dx^{\nu}  \\
{\cal A}_{(2)}  \\
\fft12 {\td b}_{\mu\nu}\,dx^{\mu}\wedge dx^{\nu}
\end{array}\right).
\eea
By comparing (\ref{gt}) with (\ref{g't}), one can read off all the
deformed fields. By reading off the deformed fields from (A.15) of
\cite{LM}, we find that the resulting solution can be written
(with the metric expressed in the string frame) as
\bea ds_{IIB}^2 &=& G\sqrt{H}\, F \Big[
\fft{1}{\sqrt{\Delta}}(D\varphi_1-C\, D\varphi_2)^2 +\sqrt{\Delta}
D{\varphi_2}^2\Big] + \sqrt{H}\, F^{-1/3}\, g_{\mu \nu} dx^{\mu}
dx^{\nu}\,,
\nonu \\
B & = & \gamma F^2 G\, D \varphi_1 \wedge D \varphi_2 -\sigma\,
{\cal A}_{(2)}\,, \qquad C^{(2)}  =  -\sigma F^2 G\, D \varphi_1
\wedge D \varphi_2 -\gamma\, {\cal A}_{(2)}\,,
\nonu \\
e^{2\phi} & = & H^2 G\,,\qquad \chi  =  \gamma \sigma F^2
H^{-1}\,, \label{deformed} \eea
and
\be F_{(5)} = dC_{(4)}+\ast\, dC_{(4)}\,,\qquad C_{(4)} = \Big(
f^{-1} \fft{r^4}{R^4} -1\Big) \,dt\wedge dx_1\wedge dx_2\wedge
dx_3\,, \ee
where the metric functions are given by (\ref{metricfunction}),
the covariant derivatives are defined in (\ref{covariant}), and $d
{\cal A}_{(2)}$ is given by (\ref{da2}). Note that the Hodge dual
in $F_{(5)}$ is now taken with respect to the deformed metric.

As can be seen from the above solution, the $\beta$-deformation
results in turning on the complex 3-form field strength, which is
given by $d \left(B + i C_{(2)}\right) = \beta\, d \left( F^2 G D
\varphi_1 \wedge D \varphi_2 -i {\cal A}_{(2)} \right)$. This
supports D5 and NS5-branes. As discussed in \cite{LM}, in order
for the 5-brane charges to be quantized, $\beta$ should take on
fractional values. Also, since the dilaton is no longer constant,
the corresponding super Yang-Mills coupling constant runs along
the flow.

One can check that the deformed geometry is regular, other than at
the singular shell of D3-branes. In fact, it is easy to see that
the deformations do not result in additional power-law type
curvature singularities to the geometry, since there are no new
poles in the metric. This is a general statement that applies to
this choice of $SL(3,R)$ deformations \cite{LM}, including all of
the examples discussed in this paper. However, since the initial
solution has poles in the metric, one has to check that the
deformations do not result in conical singularities. For the
present solution, the metric components either blow up or vanish
when either $s_{\alpha}\rightarrow 0$ or $c_{\alpha}\rightarrow
0$, $s_{2\theta}\rightarrow 0$. In either case $F$ vanishes, which
means that both $G$ and $H$ go to unity. This guarantees that the
deformed metric reduces to the original metric near the poles, so
that the regularity conditions on coordinate periods are
unchanged. We have also checked that all of the fields are
single-valued at these poles. These conclusions also apply to the
other examples to be discussed.

\section{Supersymmetric deformations of
twisted field\\ theories}

We will now consider deformations of field theories which live on
D3-branes wrapped on $H^2$, where $H^2$ is a (quotiented)
hyperbolic 2-cycle \cite{MN1}. These theories are coupled to an
$SO(6)$ gauge field and are known as twisted field theories
\cite{twisted}. In some cases, the corresponding geometry in type
IIB supergravity exhibits a smooth flow from an asymptotically
locally $AdS_5\times S^5$ geometry to a warped product of
$AdS_3\times H^2$ and an internal space. This corresponds to an RG
flow ``across dimensions'' from a four-dimensional superconformal
field theory to a two-dimensional superconformal field theory. The
2-cycle can be embedded within the rest of the internal space in a
number of different ways, which corresponds to a certain choice of
external $SO(6)$ gauge fields on the worldvolume theory. This
dictates the amount of supersymmetry preserved by the
two-dimensional theory in the IR. We will consider a general form
of the solution as it was presented in \cite{MN1} which
encompasses the cases in which the two-dimensional field theory
preserves either $(4,4)$ or $(2,2)$ supersymmetry.

First, we will turn to some of the details of the system in
five-dimensional gauged supergravity, which involves the metric, a
$U(1)$ gauge field and a scalar field which corresponds to turning
on an operator in the {\bf 20} of $SO(6)$. Consider an ans\"{a}tz
of the form
\bea ds_5^2 &=& e^{2f(r)} (dr^2+dz^2-dt^2)+\fft{e^{2g(r)}}{y^2}
(dx^2+dy^2)\,,\nnn\\
A_{(1)}^1 &=& A_{(1)}^2=a\,\fft{dx}{y}\,,\qquad
A_{(1)}^3=(1-2a)\,\fft{dx}{y}\,, \label{5Dsystem} \eea
along with a scalar field $\varphi$. The supersymmetry variation
equations of the fermionic fields imply that $f$, $g$ and
$\varphi$ obey a system of first-order differential equations,
which are written explicitly in \cite{MN1}. In particular, it was
found that taking $a=0$ or $1/2$ leads to cases in which the
two-dimensional field theory has $(4,4)$ or $(2,2)$ supersymmetry,
respectively.

The exact solution found in \cite{MN1} with $(4,4)$ supersymmetry
actually flows to a singularity in the IR. Also, the equations
were only partially solved for the $(2,2)$ supersymmetric case
which smoothly interpolates between local $AdS_5$ and $AdS_3\times
H^2$. Smooth interpolations between these conformal fixed points
have been found numerically \cite{CLV}. Solutions which
interpolate between local $AdS_5$ and $AdS_3\times S^2$ have also
been found, although the dual gauge theory has not been discussed.
For the case in which the three $U(1)$ gauge fields are all
equally-charged and the scalar is constant, there is a exact
solution \cite{sabra,KS}. However, since some fields acquire
fractional spins, it is not clear if this solution actually makes
any sense \cite{MN1}. Nevertheless, since deformations of the
types of solutions discussed in this section have similar form to
this exact solution, we provide details of this in section 4.4.

The system described by (\ref{5Dsystem}) can be lifted to type IIB
theory using the dimensional reduction ans\"{a}tz in
\cite{tenauthors}. We can express the resulting ten-dimensional
metric in $T^2$ form as\footnote{Note that $R=g^{-1}$, where $g$
is the cosmological parameter written in \cite{tenauthors}.}
\bea ds_{10}^2 &=& \sqrt{{\td\Delta}} ds_5^2+F \Big[
\fft{1}{\sqrt{\Delta}} (D\varphi_1-C\,D\varphi_2)^2
+\sqrt{\Delta}\,D\varphi_2^2\Big]\\
&+& \fft{R^2}{\sqrt{{\td\Delta}}} e^{-\varphi} \Big[
(c_{\alpha}^2+e^{3\varphi}s_{\alpha}^2)\,d\alpha^2
+s_{\alpha}^2\,d\theta^2 +\fft{9}{16}k\,e^{3\varphi} s_{2\alpha}^2
s_{2\theta}^2 \Big( d\psi+\fft{1}{3R} \fft{dx}{y}\Big)^2 \Big]\,,
\eea
where the various metric functions are
\bea {\td\Delta}(r,\alpha) &=& e^{-\varphi}
s_{\alpha}^2+e^{2\varphi} c_{\alpha}^2\,,\qquad
k(r,\theta,\alpha)^{-1}=e^{3\varphi} c_{\alpha}^2+s_{\alpha}^2
c_{\theta}^2
s_{\theta}^2\,,\nnn\\
\Delta(r,\theta,\alpha)^{-1} &=& k\,s_{\alpha}^2\,,\qquad
F(r,\theta,\alpha)=\fft{R^2}{\sqrt{k\,{\td\Delta}}} e^{-\varphi}
s_{\alpha}\,,\qquad C(\theta)=-c_{\theta}^2\,, \eea
and the connection one-forms are
\bea D\varphi_i &=& d\varphi_i +A_{\psi}^i \Big(
d\psi+\fft{a_i}{R}\,\fft{dx}{y}\Big)\,,\nnn\\
A_{\psi}^1 &=& c_{2\theta}-c_{\theta}^2 A_{\psi}^2\,,\qquad
A_{\psi}^2= (2s_{\alpha}^2 s_{\theta}^2 c_{\theta}^2-
e^{3\varphi} c_{\alpha}^2)k\,,\nnn\\
a_1 &=& \fft{a\,c_{2\theta}-a_2 c_{\theta}^2
A_{\psi}^2}{A_{\psi}^1}\,,\qquad a_2= \fft{2a-k\,e^{3\varphi}
c_{\alpha}^2}{A_{\psi}^2}\,. \eea
The self-dual five-form field strength is given by
\bea F_{(5)} &=& -\fft{2}{R} [e^{-2\varphi} s_{\alpha}^2+
e^{\varphi} (1+c_{\alpha}^2)] \fft{e^{3f+2g}}{y^2}\,dr\wedge
dz\wedge dt\wedge dx\wedge dy\nnn\\ &+& \fft32 R
\fft{e^{2g+f}}{y^2} (\partial_r \varphi)\, s_{2\alpha}\, dz\wedge
dt\wedge dx\wedge dy\wedge d\alpha\nnn\eea \bea &+& \fft{R^2}{2}
e^{3f-2g} \Big[ a\,e^{2\varphi} (s_{2\alpha} c_{\theta}^2\,d\alpha
-s_{\alpha}^2 s_{2\theta}\,d\theta)\wedge \Big( d\psi+d\varphi_1
+d\varphi_2 +\fft{a}{R}\fft{dx}{y}\Big)\nnn\\ &+& a\,e^{2\varphi}
(s_{2\alpha} s_{\theta}^2\,d\alpha +s_{\alpha}^2
s_{2\theta}\,d\theta)\wedge \Big( d\psi-d\varphi_1
+\fft{a}{R}\fft{dx}{y}\Big) \label{initialtwistedF5} \\ &+& (2a-1)
e^{-4\varphi} s_{2\alpha}\,d\alpha \wedge \Big( d\psi-d\varphi_2
+\fft{1-2a}{R}\fft{dx}{y}\Big) \Big]\wedge dr\wedge dz\wedge dt
+{\rm dual}\,.\nnn \eea

Computing the Hodge dual terms in $F_{(5)}$ explicitly gives the
term that is relevant for computing contributions to the $C^{(2)}$
and $B$ fields in the deformed solution. This term is given by
\be F_{(5)}=d{\cal A}_{(2)}\wedge D\varphi_1\wedge
D\varphi_2+\cdots\,, \label{A2} \ee
where $d{\cal A}_{(2)}$ has the form
\bea d{\cal A}_{(2)} &=& \Big( F_{\alpha\theta\psi}\,
d\alpha\wedge d\theta + F_{r\theta\psi} \,dr\wedge d\theta
+F_{y\theta\psi}\,dy\wedge d\theta+a\,F_{y\alpha\psi}\,dy\wedge
d\alpha\Big) \wedge \Big( d\psi+\fft{1}{3R}\fft{dx}{y}\Big)\nnn\\
&+& \Big( F_{xy\theta}\, d\theta +a\, F_{xy\alpha}\, d\alpha\Big)
\wedge dx\wedge dy\,. \eea
The $F_{\mu\nu\lambda}$ coefficients are presented in their full
glory in appendix A. In the above expression for $F_{(5)}$ and for
the rest of the paper, the $\cdots$ represent terms which do not
include $D\varphi_1\wedge D\varphi_2$ in the wedge product. These
terms do not contribute to the final two-form fields, as can be
seen from the deformation formulae in \cite{LM}.

The resulting deformed solution is given by the expression
(\ref{deformed}) along with the 5-form in
(\ref{initialtwistedF5}). The 5-form is the same as before the
deformation except that the Hodge dual is taken with respect to
the deformed metric.

\section{Further examples of supersymmetric deformations}

We will now consider various backgrounds in type IIB theory which
are asymptotically locally (and in some cases globally)
$AdS_5\times S^5$. On general grounds, one expects that these
gravity solutions are dual to four-dimensional super Yang-Mills
theories which exhibit RG flows from an ${\cal N}=4$
superconformal fixed point in the UV limit. The deformed theories
have an ${\cal N}=1$ superconformal fixed point in the UV limit.
In all of these examples, the dual field theory lives on a curved
background. In sections 4.1 and 4.2, the dual field theory lives
on $\R\times S^3$. In sections 4.3 and 4.4, the field theory is on
Minkowski$_2\times H^2$ (like the twisted theories of section 3)
and $AdS_3\times S^1$, respectively.

\subsection{Supersymmetric ${\rm AdS}_5$ black hole}

As we already mentioned, the maximally-supersymmetric continuous
distributions of D3-branes discussed in section 2 tend to be
singular along the surface containing the branes. This means that
the corresponding solution in five-dimensional gauged supergravity
is actually a naked singularity, rather than a black hole. On the
other hand, rotating black D3-branes do correspond to black holes
in five dimensions, but these are not supersymmetric. It has only
been quite recently that supersymmetric $AdS_5$ black holes have
been known \cite{reall1}. It is a requirement that there is
nonvanishing angular momentum. These solutions were coupled to
arbitrarily many abelian vector multiplets in \cite{reall2}.

For the case in which there is a single gauge field\footnote{We
absorbed a $1/R$ factor in the gauge field for typographical
convenience.}, the supersymmetric $AdS_5$ black hole solution is
given \cite{reall1}
\bea ds_5^2 &=& -{\cal F}^2 (dt+\Psi\,\sigma_3)^2+
\fft{R^2\,dr^2}{{\cal F}^2 (R^2+2\omega^2+r^2)} +\fft{r^2}{4}
\Big[ \sigma_1^2+\sigma_2^2+\Big(
\fft{R^2+2\omega^2+r^2}{R^2}\Big)\, \sigma_3^2\Big]\,,\nnn\\
A_{(1)} &=& \frac{1}{R} \Big[ {\cal F}\,dt +\fft{\eta
\omega^4}{4Rr^2}\,\sigma_3\Big] \,, \label{5metric1} \eea
where
\be {\cal F}=1-\fft{\omega^2}{r^2}\,,\qquad \Psi=-\fft{\eta
r^2}{2R}\,\Big( 1+\fft{2\omega^2}{r^2}+
\fft{3\omega^4}{2r^2(r^2-\omega^2)}\Big) \,. \ee
$\eta=\pm 1$ and $\omega$ is a constant. The $\sigma_i$ are
left-invariant $SU(2)$ one-forms given by
\bea \sigma_1 &=& s_{\phi_3}\,d\phi_1-s_{\phi_1}\,c_{\phi_3}\,
d\phi_2\,,\quad \sigma_2 =
c_{\phi_3}\,d\phi_1+s_{\phi_1}\,s_{\phi_3}\,d\phi_2\,, \quad
\sigma_3 = d\phi_3+c_{\phi_1}\,d\phi_2\,. \label{sigma} \eea
It has been shown that the above geometry is globally $AdS_5$ at
large distance \cite{reall1}.

This solution can be lifted to ten-dimensional type IIB theory
using the ans\"{a}tz in \cite{tenauthors}, which yields
\bea ds_{10}^2 &=& ds_5^2 + R^2\, [d\alpha^2+s_{\alpha}^2
d\theta^2+ s_{\alpha}^2 c_{\theta}^2
(d\psi+d\varphi_1+d\varphi_2+ A_{(1)})^2\nnn\\
&+& c_{\alpha}^2 (d\psi-d\varphi_2+ A_{(1)})^2 +s_{\alpha}^2
s_{\theta}^2 (d\psi-d\varphi_1+ A_{(1)})^2 ]\,,\nnn\\
F_{(5)} &=& \fft{r^3}{2R}\,dt\wedge dr\wedge \sigma_1\wedge
\sigma_2\wedge \sigma_3 + \fft{\omega^2}{8r^3} [d(\mu_1^2)\wedge
(d\psi-d\varphi_2+ A_{(1)})\nnn\\ &+& d(\mu_3^2)\wedge
(d\psi-d\varphi_1+ A_{(1)}) + d(\mu_2^2)\wedge
(d\psi+d\varphi_1+d\varphi_2+ A_{(1)})]\nnn\\ &\wedge& [ r^3
(R^2+2\omega^2+r^2)\, \sigma_1\wedge \sigma_2\wedge\sigma_3 +R
{\cal F}^2 r\,( 4R \Psi+\eta \omega^2)\,(dt+\Psi\,\sigma_3) \wedge
\sigma_1\wedge \sigma_2\nnn\\ &+& 2R \eta \omega^2\, dt\wedge
dr\wedge \sigma_3 ] +{\rm dual} \,, \label{metricandfive} \eea
where
\be \mu_1=c_{\alpha}\,\qquad \mu_2=s_{\alpha}\,c_{\theta}\,,\qquad
\mu_3=s_{\alpha}\,s_{\theta}\,. \label{mu} \ee
It has been shown that this solution preserves two supersymmetries
in type IIB theory \cite{gauntlett1} At large distance, this
solution is dual to an ${\cal N}=4$ superconformal field theory.
It is not known if this complete solution can arise in the
near-horizon limit of an asymptotically flat D3-brane
configuration. However, if it does, then the D3-branes would
exhibit rotation in both the transverse space as well as on the
worldvolume.

The metric (\ref{metricandfive}) can be expressed as
\bea ds_{10}^2 &=& F\Big[ \fft{1}{\sqrt{\Delta}}(D\varphi_1-C
D\varphi_2)^2 +\sqrt{\Delta} D\varphi_2^2\Big] + ds_5^2
\nnn\\
&+& R^2 \left[ d \alpha^2 + s_{\alpha}^2 d\theta^2 + \left(
\frac{9 c_{\alpha}^2 s_{\alpha}^2 s_{2\theta}^2} {4 c_{\alpha}^2 +
s_{\alpha}^2 s_{2\theta}^2} \right) \left(d \psi +
A_{(1)}\right)^2 \right], \label{commonmetric} \eea
where the 5-metric is given by (\ref{5metric1}) and the metric
functions are given by
\bea
 F(\theta,\alpha) & = & R^2 s_{\alpha} \sqrt{c_{\alpha}^2+
s_{\alpha}^2 s_{\theta}^2 c_{\theta}^2}\,, \qquad
 \Delta(\theta,\alpha) =
\fft{c_{\alpha}^2+ s_{\alpha}^2 s_{\theta}^2 c_{\theta}^2}{
s_{\alpha}^2 }\,,\qquad C(\theta)=- c_{\theta}^2\,. \eea
We have introduced the connection one-forms
\bea D\varphi_1 & \equiv & d\varphi_1+A_{\psi}^1 \left( d\psi
+A_{(1)} \right) \,,\qquad D\varphi_2 \equiv d\varphi_2+A_{\psi}^2
\left(
d\psi +  A_{(1)} \right)\,,\nnn\\
A_{\psi}^1 &=&c_{2\theta} - c_{\theta}^2  \frac{-c_{\alpha}^2 +
2s_{\alpha}^2 s_{\theta}^2 c_{\theta}^2}{c_{\alpha}^2
+s_{\alpha}^2 s_{\theta}^2 c_{\theta}^2} \,,\qquad A_{\psi}^2 =
\frac{-c_{\alpha}^2 + 2s_{\alpha}^2 s_{\theta}^2
c_{\theta}^2}{c_{\alpha}^2 +s_{\alpha}^2 s_{\theta}^2
c_{\theta}^2}\,. \label{oneforms} \eea

Computing the Hodge dual terms in $F_{(5)}$ explicitly gives the
term of the form (\ref{A2}) which is relevant for computing the
deformed two-form fields, where
\bea d{\cal A}_{(2)} &=& 6R^4\,s_{2\theta} s_{\alpha}^3
c_{\alpha}\, d\alpha\wedge d\theta \wedge (d\psi+A_{(1)})\nnn\\
&+& \fft{\omega^2 R^2}{48r^3}(4c_{\alpha}^2+s_{\alpha}^2
s_{2\theta}^2)\Big( \fft{2s_{\alpha}^2}{s_{2\theta}}
[(1+c_{\theta}^2)A_{\psi}^2+c_{2\theta}A_{\psi}^1]
+\fft{s_{\alpha}}{c_{\alpha}}(2A_{\psi}^1+A_{\psi}^2)\Big)\\
&\times& [8R\, (dt+\Psi\,\sigma_3)\wedge dr\wedge d\alpha+2
(4R\Psi+\eta\omega^2)\, dr\wedge \sigma_3\wedge d\alpha
+\eta\omega^2 r\,\sigma_1\wedge \sigma_2\wedge d\alpha \,.\nnn
\eea
The resulting deformed solution is given by the expression
(\ref{deformed}) along with the 5-form in (\ref{metricandfive}).
The 5-form is the same as before the deformation except that the
Hodge dual is taken with respect to the deformed metric.

\subsection{Deformation of $AdS_5$}

A one-parameter family of type IIB solutions that is
asymptotically locally $AdS_5\times S^5$ was found in
\cite{gauntlett,gauntlett1}. This is referred to as deformation of
$AdS_5\times S^5$ by the authors, which is a different type of
deformation than the one we are presently implementing. The metric
has the form (\ref{metricandfive}) with $ds_5^2$ given by
\be ds_5^2 = -\Big(
dt+\fft{r^2}{2R}\sigma_3+\fft{fr^2}{V}\sigma_1\Big)^2+
\fft{dr^2}{V}+\fft{r^2}{4}(\sigma_1^2+\sigma_2^2+V\,\sigma_3^2)\,,
\label{deformed5metric} \ee
where
\be V(r)=1+r^2/R^2\,,\qquad A_{(1)}=\fft{fr^2}{R\,V(r)}\,
\sigma_1\,, \label{functions} \ee
and the $\sigma_i$ and $\mu_i$ are given by (\ref{sigma}) and
(\ref{mu}), respectively. There is also a self-dual five-form
field given by
\bea F_{(5)} &=& \fft{r^3}{2R}\,dt\wedge dr\wedge \sigma_1\wedge
\sigma_2\wedge\sigma_3-\fft{f\,R^2}{2V^2} [d(\mu_1^2)\wedge
(d\psi-d\varphi_2+A_{(1)}) \nnn\\
&+& d(\mu_3^2)\wedge
(d\psi-d\varphi_1+A_{(1)}) + d(\mu_2^2)\wedge
(d\psi+d\varphi_1+d\varphi_2+A_{(1)})]\label{deformed5form1} \\
&\wedge& \Big[dt\wedge \Big(r^2V\,\sigma_2\wedge\sigma_3-2r\,
dr\wedge\sigma_1\Big) - \sigma_1\wedge \sigma_3\wedge \Big(
\fft{r^3}{R}\,dr+f\,r^4\,\sigma_2\Big) \Big]+{\rm dual} \,. \nnn
\eea
This solution preserves two supersymmetries of type IIB theory
\cite{gauntlett1}. When the parameter $f$ vanishes, the above
geometry reduces to $AdS_5\times S^5$. For nonzero $f$, the
geometry is locally asymptotically $AdS_5\times S^5$, implying
that the UV limit of the dual gauge theory is ${\cal N}=4$ super
Yang-Mills theory.

This geometry has the exotic property that, when $4R^2 f^2>1$,
closed timelike curves are present; they are absent when $4R^2
f^2\le 1$ \cite{gauntlett}. These closed timelike curves appear in
the boundary, as well as the bulk, which implies that they are
present on the background of the dual gauge theory. It has been
shown that the holographic energy-momentum tensor remains finite,
which might be interpreted as a hint that the AdS/CFT
correspondence is still valid in the presence of closed timelike
curves. Of course, it is certainly possible that these closed
timelike curves are simply a sign that the solution is not
physical past an upper bound on $f$ \cite{gauntlett1}.

The 10-dimensional metric can be written in the form
(\ref{commonmetric}) together with (\ref{deformed5metric}) and
(\ref{functions}). Computing the Hodge dual terms in $F_{(5)}$
explicitly gives the relevant term (\ref{A2}) with
\bea d{\cal A}_{(2)} &=& 6R^4\,s_{2\theta} s_{\alpha}^3
c_{\alpha}\, d\alpha\wedge d\theta \wedge (d\psi+A_{(1)})\\
&-& \fft{fR^3r}{6V^2} (4c_{\alpha}^2+s_{\alpha}^2 s_{2\theta}^2)
(2dr\wedge \sigma_1-rV\,\sigma_2\wedge \sigma_3)\nnn\\ &\wedge&
\Big[ \fft{s_{\alpha}^3}{s_{2\theta}}[(1+c_{\theta}^2)
A_{\psi}^2+c_{2\theta} A_{\psi}^1]\,d\theta
+\fft{s_{\alpha}^2}{2c_{\alpha}}
(2A_{\psi}^1+A_{\psi}^2)\,d\alpha\Big] \,.\nnn \eea
The resulting deformed solution is given by the expression
(\ref{deformed}) along with the 5-form in (\ref{deformed5form1}).

\subsection{$AdS_5$ magnetic string}

We will now discuss the deformations of a five-dimensional
magnetic string solution which can be embedded in ${\cal N}=2$
gauged supergravity\footnote{It has been shown that this solution
solves first-order equations which come from a superpotential
\cite{CLV}.} \cite{sabra,KS}. This solution arises from the system
studied in section 3 when the three $U(1)$ gauge fields are all
equally-charged and the scalars are constant. However, as
previously mentioned, in this case some fields acquire fractional
spins, which might be a sign that this solution is not
well-behaved \cite{MN1}. Nevertheless, it is a simple and exact
solution which shares some of the properties of other solutions
which arise from the system discussed in section 3. The
five-dimensional solution is given by
\bea ds_5^2 &=& \fft{r^2}{R^2} H^{3/2}
(-dt^2+dx^2)+\fft{R^2}{H^2r^2} dr^2+r^2
(d\phi_1^2+\sinh^2\phi_1\,d\phi_2^2)\,,\nnn\\
F_{(2)} &=& -\fft{R}{3}\,\sinh\phi_1\,d\phi_1\wedge d\phi_2\,,
\label{fivemetric} \eea
where
\be H=1-\fft{R^2}{3r^2}\,. \ee
The geometry interpolates between $AdS_3\times H^2$, where $H^2$
is hyperbolic 2-plane, at short distance and locally $AdS_5$ at
large distance. At short distance, the supersymmetry is enhanced
by a factor of two \cite{KS}. The charge of the $U(1)$ gauge field
is uniquely determined by a supersymmetry condition.

This solution can be lifted to type IIB theory, where it preserves
two supersymmetries. Using the dimensional reduction ans\"{a}tz
given in \cite{tenauthors}, one can see that the ten-dimensional
metric has the form (\ref{metricandfive}), where $ds_5^2$ is given
in (\ref{fivemetric}) and $A_{(1)}=\ft13 \cosh\phi_1\,d\phi_2$.
The self-dual five-form is given by
\bea F_{(5)} &=& \fft{4r^3\sqrt{H}}{R^2}\sinh\phi_1\,dt\wedge
dx\wedge
dr\wedge d\phi_1\wedge d\phi_2\nnn\\
&+& \fft{\sqrt{H}R^2}{6r}\,dt\wedge dx\wedge dr\wedge
[d(\mu_1^2)\wedge (d\psi-d\varphi_2+ A_{(1)}) \label{initialtwisted} \\
&+& d(\mu_3^2)\wedge (d\psi-d\varphi_1+ A_{(1)})+ d(\mu_2^2)\wedge
(d\psi+d\varphi_1+d\varphi_2+ A_{(1)})]+{\rm dual} \,.\nnn \eea
where we take the Hodge dual with respect to the five-dimensional
metric (\ref{fivemetric}). Then the ten-dimensional metric can be
written in the form (\ref{commonmetric}).

Computing the Hodge dual terms in $F_{(5)}$ explicitly gives
(\ref{A2}) where
\bea d{\cal A}_{(2)} &=& 6R^4\,s_{2\theta} s_{\alpha}^3
c_{\alpha}\, d\alpha\wedge d\theta \wedge (d\psi+A_{(1)}) -
\fft{R^4}{6}(4c_{\alpha}^2+s_{\alpha}^2 s_{2\theta}^2)\\
&\times& \Big[ \fft{s_{\alpha}}{c_{\alpha}}
(2A_{\psi}^1+A_{\psi}^2)\, d\alpha -
\fft{s_{\alpha^2}}{s_{2\theta}} [c_{2\theta}
A_{\psi}^2+(1+c_{\theta}^2)A_{\psi}^2]\,
d\theta\Big]\sinh\phi_1\,d\phi_1\wedge d\phi_2 \,.\nnn \eea
The resulting deformed solution is given by the expression
(\ref{deformed}) along with the 5-form in (\ref{initialtwisted}).
Like the solutions discussed in section 3, this may be dual to an
four-dimensional ${\cal N}=1$ superconformal field theory in the
UV limit which flows to a two-dimensional $(2,2)$ superconformal
field theory in the IR region.

\subsection{$AdS_5$ flux-brane/anti-bubble}

The magnetic string geometry of the previous subsection and some
of the five-dimensional geometries discussed in section 3 are
locally $AdS_5$ in the UV limit and $AdS_3\times H^2$ in the IR
region. In fact, the same $AdS_3\times H^2$ fixed point can
interpolate to two different asymptotically locally $AdS_5$
spacetimes. For the previously-discussed case, the local $AdS_5$
geometry has a boundary Minkowski$_2\times H^2$. This is because
there is an $H^2$ component of the metric, with a warping factor,
throughout the flow. It is also possible to maintain an $AdS_3$
component, again with a warping factor, all along the flow. In
this case, the warp factor depends on a ``radial'' coordinate
which lies within the $H^2$ rather than $AdS_3$, and the
asymptotically locally $AdS_5$ has the boundary $AdS_3\times S^1$.
This five-dimensional solution \cite{LPV} is given by
\bea ds_5^2 = H\,dx^2+H^{-1}dr^2+r^2\,ds_{{\rm AdS}_3}^2\,,\qquad
\ast_5 F_{(2)} = \fft{8}{9\sqrt{3}}R^2\,\epsilon_{(3)}\,,
\label{metricfive} \eea
where
\be H=\fft{r^2}{R^2}-\Big( 1-\fft{4R^2}{27r^2}\Big)^2\, \ee
and $\epsilon_{(3)}$ is the volume-form of the unit ${\rm AdS}_3$.
This solution is related to the supersymmetric $AdS_5$ naked
singularity discussed in section 2 via a double Wick rotation.
This solution smoothly flows between the asymptotically locally
$AdS_5$ region and $AdS_3\times H^2$ at $r=\ft23 R$. This was
initially called a flux-brane because the supporting magnetic
two-form field strength $F_{(2)}$ is proportional to the entire
transverse space spanned by the coordinates $r$ and $x$
\cite{LPV}. More recently, it has been referred to as an
anti-bubble \cite{jones}. It is interesting to note that, once the
cosmological constant is fixed, there are no free parameters
associated with the $F_{(2)}$ field strength or the $AdS_3$
curvature. Equivalently, the $AdS_3$ curvature fixes the
five-dimensional cosmological constant as well as the charge
associated with $F_{(2)}$.

Lifting this solution to type IIB theory with the ans\"{a}tz given
in \cite{tenauthors} yields a ten-dimensional metric of the form
(\ref{metricandfive}) with $ds_5^2$ given by (\ref{metricfive})
and $A_{(1)}=\ft{R}{2r^2}\,dx$. The self-dual five-form is given
by
\bea F_{(5)} &=& \fft{4r^3}{R}\,\epsilon_{(3)}\wedge dx\wedge dr
 - \fft{4}{27}R^4\,\epsilon_{(3)}\wedge \Big[
d(\mu_1^2)\wedge \Big(
d\psi-d\varphi_2+A_{(1)}\Big) \label{F5fluxbrane} \\
&+& d(\mu_3^2)\wedge \Big( d\psi-d\varphi_1+A_{(1)}\Big) +
d(\mu_2^2)\wedge \Big( d\psi+d\varphi_1+d\varphi_2+A_{(1)}\Big)
\Big] +{\rm dual} \,,\nnn \eea
where the $\mu_i$ are given by (\ref{mu}). Then the 10-dimensional
metric can be written in the form (\ref{commonmetric}) with
$D\varphi_i$ and $A_{\psi}^i$ given by (\ref{oneforms}). This
solution might be related to some of those recently discussed in
\cite{lin}.

The gravity background asymptotically has $SO(2,4)\times SO(6)$
symmetry, where $SO(2,4)$ describes the four-dimensional conformal
symmetry of the UV fixed point with $SO(6)$ R-symmetry. Away from
the UV limit, the symmetry of the gravity background is broken to
$SO(2,2)\times U(1)^4$. On the gauge theory side, this corresponds
to the four-dimensional conformal symmetry breaking to $SO(2,2)$,
which is the $SL(2,R)\times SL(2,R)$ part of the three-dimensional
conformal symmetry group. Also, a $U(1)^3$ of the R-symmetry group
remains unbroken. There is an additional $U(1)$ corresponding to a
circular direction in the background on which the field theory
lives. As the IR region is approached, the four-dimensional
description breaks down and is replaced by a three-dimensional
description. Therefore, this gravity solution probably corresponds
to a vacuum of ${\cal N}=4$ Yang-Mills theory on $AdS_3\times
S^1$, for which a Kaluza-Klein mode on $S^1$ is
excited\footnote{We appreciate Juan Maldacena's input on this
matter.}. This KK mode corresponds to the gauge field in
(\ref{metricfive}) from the point of view of five-dimensional
gauged supergravity. From the ten-dimensional type IIB point of
view, this corresponds to gravitational fluctuations associated
with the $U(1)^3$ directions of the five-sphere. The scale of the
KK mode is given by the parameter $R$. Also, this classical
gravity description is valid so long as $R$ is large compared to
the string length.

The term in $F_{(5)}$ which is relevant for the deformation is
(\ref{A2}) where
\bea d{\cal A}_{(2)} &=& 6R^4\,s_{2\theta} s_{\alpha}^3
c_{\alpha}\, d\alpha\wedge d\theta \wedge (d\psi+A_{(1)})\\
&-& \fft{2R^5}{81r^3}(4c_{\alpha}^2+s_{\alpha}^2 s_{2\theta}^2)
\Big[ \fft{2s_{\alpha}^2}{s_{2\theta}}[c_{2\theta}A_{\psi}^1
+(1+c_{\theta}^2)A_{\psi}^2]\,d\theta-\fft{s_{\alpha}}{c_{\alpha}}
(2A_{\psi}^1+A_{\psi}^2)\,d\alpha\Big]\wedge dx\wedge dr \,.\nnn
\eea
The resulting deformed solution is given by the expression
(\ref{deformed}) along with the 5-form in (\ref{F5fluxbrane}). The
deformed dual field theory is a four-dimensional ${\cal N}=1$
superconformal field theory on $AdS_3\times S^1$ in the UV limit
and flows to a two-dimensional $(2,2)$ superconformal field theory
in the IR region.

\section{Nonsupersymmetric deformations}

We now turn to a couple of examples of deformations which do not
preserve any supersymmetry. Both of the initial solutions are
generalizations of the ${\cal N}=4$ Coulomb branch flow discussed
in section 2. In section 5.1, we discuss an ${\cal N}=1$
supersymmetric generalization given in \cite{KW,GW}. Although the
initial solution is supersymmetric, the resulting solution is
nonsupersymmetric since the deformation involves the direction
associated with the $U(1)$ R-symmetry. In section 5.2, we consider
a case in which the initial solution itself is nonsupersymmetric.
In particular, we consider the near-horizon limit of a nonextremal
rotating D3-brane, which corresponds to the gauge theory being at
finite temperature \cite{larsen}.

\subsection{${\cal N}=1$ Pilch-Warner flow}

There is a three-parameter family of type IIB supergravity
solutions which correspond to\newline ${\cal N}=1$ supersymmetric
holographic RG flows \cite{KW,GW}. Two of the parameters represent
independent scalar masses or vevs and the third represents a
single fermion mass. The ten-dimensional metric of the full
solution is given by \cite{KW,GW}
\bea ds_{IIB}^2 = \Omega^2 \left( e^{2A(r)}dx_{\mu}^2 + dr^2
\right) + \fft{L^2}{\Omega^2}\,ds_5^2 \label{10dmetric} \eea
where the warp factor is
\bea \Omega(r,\theta,\phi) \equiv \sqrt{ \cosh \chi}
\left[\frac{1}{\rho^2} \left( \nu^2 c_{\phi}^2 + \frac{1}{\nu^2}
s_{\phi}^2 \right) c_{\theta}^2  + \rho^4 s_{\theta}^2
\right]^{1/4}\,, \eea
and the metric of the five-dimensional internal space is given by
\bea ds_5^2 & = & \frac{1}{\rho^4}[c_{\theta}^2 +\rho^6
s_{\theta}^2 (\nu^{-2} c_{\phi}^2 +\nu^2 s_{\phi}^2)]\, d\theta^2
+ \rho^2 c_{\theta}^2 [\nu^2 c_{\phi}^2 +\nu^{-2} s_{\phi}^2]\,
d\phi^2\nnn\\ &-& 2\rho^2 [\nu^2-\nu^{-2}] s_{\theta} c_{\theta}
s_{\phi} c_{\phi}\, d\theta d \phi + \rho^2 c_{\theta}^2 [\nu^{-2}
c_{\phi}^2 d \varphi_1^2 + \nu^2 s_{\phi}^2 d \varphi_2^2] +
\frac{1}{\rho^4}
s_{\theta}^2 (d\varphi_1+d\varphi_2+d\varphi_3)^2 \nonu \\
&+ & \frac{1}{\Omega^4} \sinh^2 \chi \cosh^2 \chi\, [c_{\theta}^2
(s_{\phi}^2\, d \varphi_2-c_{\phi}^2\, d \varphi_1)+s_{\theta}^2(
d\varphi_1+d\varphi_2+d\varphi_3)]^2\,. \label{5metric} \eea
From the five-dimensional gauged supergravity perspective, the
three scalar fields $\chi(r),\nu(r)$ and $\rho(r)$ satisfy
first-order equations which arise from a superpotential. The
fermion mass vanishes asymptotically when $\chi=0$, where the
solution reduces to that of the near-horizon region of an
ellipsoidal distribution of extremal D3-branes (\ref{KLT}). This
limit, which we have already discussed in section 2, is associated
with the Coulomb branch flows of the ${\cal N}=4$ super Yang-Mills
theory. When $\nu=1$, this system reduces to the Pilch-Warner
flows \cite{PWfeb2000,PW1}, which correspond to the
Leigh-Strassler renormalization flow studied in \cite{LS}. We will
focus on this limit, in which case the solution can be written
as\footnote{This form of the solution can be obtained by taking
$\phi\rightarrow \phi -\beta$ in the solution of \cite{PW1} in
order to absorb the $\beta$ dependence on two-form potential. The
$U(1)_{\beta}$ symmetry under which $\Phi_1, \Phi_2$ and $\Phi_3$
have charges $(1/2,1/2,0)$ exists if we allow the mass $m$ of the
chiral superfield $\Phi_3$ to rotate by a phase \cite{GW}: $
\Phi_1 \rightarrow e^{\frac{i }{2} \beta} \Phi_1, \Phi_2
\rightarrow e^{\frac{i }{2} \beta } \Phi_2, \Phi_3 \rightarrow
\Phi_3, m \rightarrow   e^{i \beta}  m$. However, since the
original $\phi$ rotation generates $U(1)_R$ symmetry, after this
shift, still $\beta$ rotation is related to this $U(1)_R$
symmetry. In other words, contrary to the previous case, one of
the $U(1)$'s we take should contain $U(1)_R$ implying that the
supersymmetry is broken completely.}
\bea ds_{IIB}^2 &=& \fft{X_1^{1/2}\cosh\chi}{\rho}
(e^{2A(r)}dx_{\mu}^2+dr^2)\nnn\\ &+&
\fft{X_1^{1/2}\rho^3}{g^2\,{\rm cosh}\chi}\Big[ \fft{4}{\rho^6}
d\theta^2 +
\fft{c_{\theta}^2}{X_1}(d\alpha^2+s_{\alpha}^2\,d\eta^2) +
\fft{c_{\theta}^2\cosh\chi\,}{X_2}
(d\beta+c_{\alpha}\,d\eta)^2\nnn\\ &+&
\fft{s_{\theta}^2\cosh\chi\,}{X_1^2 X_2 \rho^6} \Big( 2X_2
(d\phi-d\beta)+\rho^6 c_{\theta}^2\,\sinh\chi\,\tanh\chi\, (d\beta
+c_{\alpha}\,d\eta)\Big)^2\Big]\,,\nnn\\
F_{(5)} &=& d^4 x \wedge (w_r\, dr+w_{\theta}\, d\theta)+\ast
\left[d^4 x \wedge
(w_r\, dr+w_{\theta}\, d\theta)\right] \,,\nnn\\
B_{(2)}^{{\rm NS}}+i\,C_{(2)}^{{\rm RR}} &=& e^{-i\phi}\Big[
i\,a_1d\theta-a_2 (d\beta+c_{\alpha}\,d\eta)-a_3
(d\phi-d\beta)\Big] \wedge (d\alpha +i\,s_{\alpha}\,d\eta)\,.
\label{PWmetric} \eea
The functions $X_1$, $X_2$, $w_r$, $w_{\theta}$ and $a_i$ depend
on $r$ and $\theta$ and are given by (\ref{X}), (\ref{w}) and
(\ref{a}). They are written in terms of the scalar fields
$\chi(r)$ and $\rho(r)$, which satisfy the supersymmetric flow
equations \cite{FGPW} of five-dimensional gauged supergravity.
These flow equations also involve the metric function $A(r)$,
although no explicit closed analytic solutions are known. The
dilaton-axion scalar field remains constant along this flow. The
complex two-form field corresponds to non-zero fermion masses in
the super Yang-Mills theory. The geometry has $SU(2)\times
U(1)_{\beta}$ isometry, which contains a $U(1)_{\eta}\times
U(1)_{\beta}$ subgroup. The $U(1)_{\eta} \times U(1)_{\beta}$
correspond to the $\eta$ and $\beta$ directions along which we
will be applying the $SL(2,R)$ transformations.

At large distance, the solution has the maximally supersymmetric
geometry $AdS_5\times S^5$, which is dual to ${\cal N}=4$
superconformal Yang-Mills theory. At short distance, the geometry
is a warped product of $AdS_5$ and the internal space, which
corresponds to the IR ${\cal N}=1$ superconformal fixed point of
the Leigh-Strassler flow \cite{LS}.

We will now express this solution in the form given by equation
(A.7) in \cite{LM}, in order to apply their deformation procedure:
\bea ds_{IIB}^2 &=& F\Big[ \fft{1}{\sqrt{\Delta}}(D\beta-C
D\eta)^2 +\sqrt{\Delta} D\eta^2\Big] +\fft{e^{2\phi/3}}{F^{1/3}}
\Big(\fft{X_1 \cosh\chi}{g}\Big)^{2/3} (c_1 c_4)^{1/6}\nnn\\
&\times& \Big[ e^{2A(r)} dx_{\mu}^2+dr^2+\fft{\rho^4}{g^2
\cosh^2\chi} \Big( \fft{4}{\rho^6}
d\theta^2+\fft{c_{\theta}^2}{X_1}d\alpha^2 +c_6\,
d\phi^2\Big)\Big]\,, \label{LMmetric}
\\
B_{(2)}^{{\rm NS}} &=& B_{12} D\beta\wedge D\eta+ (B_{1\mu}
D\beta+B_{2\mu}D\eta)\wedge dx^{\mu} - \fft12 A_{\mu}^m B_{m\nu}
dx^{\mu}\wedge dx^{\nu}+\fft12 {\td b}_{\mu\nu}\,
dx^{\mu}\wedge dx^{\nu}\,,\nnn\\
C_{(2)}^{{\rm RR}} &=& C_{12} D\beta\wedge D\eta+ (C_{1\mu} D\beta
+C_{2\mu} D\eta)\wedge dx^{\mu} - \fft12 A_{\mu}^m C_{m\nu}
dx^{\mu}\wedge dx^{\nu}+\fft12 {\td c}_{\mu\nu}\, dx^{\mu}\wedge
dx^{\nu}\,.
 \nonu
\eea
The index $m$ stands for 1 and 2 that are the coordinates $\beta$
and $\eta$ for two torus and the indices $\mu, \nu$ stand for the
remaining 8-dimensional coordinates and
\bea D\beta & \equiv & d\beta+A_{\phi}^1 d\phi
\,,\qquad D\eta \equiv d\eta+A_{\phi}^2 d\phi\,,\nnn\\
A_{\phi}^1 &=& (c_3-c_2 c_5)\,,\qquad A_{\phi}^2=c_5\,. \eea
Also, the metric functions are given by
\be F(r,\theta)=\fft{(X_1 c_1 c_4)^{1/2} \rho^3}{g^2
\cosh\chi}\,,\qquad \Delta(r,\theta)=\fft{c_4}{c_1}\,,\qquad
C(r,\theta)=-c_2\,. \ee
The coefficient functions in $B_{(2)}^{{\rm NS}}$ and
$C_{(2)}^{{\rm RR}}$ are given by (\ref{BC}), and the functions
$c_i$ are given in (\ref{ci}). The resulting nonsupersymmetric
type IIB deformed metric in the string frame is given by
\bea ds_{IIB}^2 &=& G\sqrt{H}\, F \Big[
\fft{1}{\sqrt{\Delta}}(D\beta'-C\,
D\eta')^2 +\sqrt{\Delta} D{\eta'}^2\Big] \nonu \\
&+& \sqrt{H}\,\fft{X_1^{1/2}\cosh\chi}{\rho} \left[ e^{2A(r)}
dx_{\mu}^2+dr^2+\fft{\rho^4}{g^2 \cosh^2\chi} \Big(
\fft{4}{\rho^6} d\theta^2+\fft{c_{\theta}^2}{X_1}d\alpha^2 +c_6\,
d\phi^2\Big)\right],
\nonu \\
D \beta' & \equiv & D \beta + (\gamma B_{2\mu}-\sigma C_{2\mu})\,
dx^{\mu} \,,\qquad D \eta'  \equiv D \eta +(\sigma C_{1\mu}-\gamma
B_{1\mu})\, dx^{\mu}\,, \eea
and the dilaton-axion scalar fields \footnote{ For nonzero
$B_{12}$ and $C_{12}$, the $SL(3,R)$ transformation yields $H
\equiv \sigma^2\,(B_{12}^2 + F^2) + (1 - \sigma\,C_{12})^2\,,\\
G^{-1} \equiv \sigma^2\,F^2 + \gamma^2\,(F^2+B_{12}^2) -
2\gamma\,B_{12}(\sigma\,C_{12}-1) + (1-\sigma\,C_{12})^2$. See
also \cite{LM,GN,Pal:2005nr}.} are
\be e^{2\phi}=H^2\,G\,,\qquad \chi =
[\gamma\sigma\,(F^2+B_{12}^2+C_{12}^2)+\sigma\,
B_{12}-\gamma\,C_{12}]H^{-1}\,, \ee
One can also straightforwardly compute the deformed form-fields by
inputting the above initial solution into the deformation formulae
in \cite{LM} but the results are too long to present. At the UV
fixed point, this deformed solution reduces to a particular
nonsupersymmetric marginal deformation of $AdS_5\times S^5$ which
was found in \cite{Frolov}.

\subsection{Nonextremal rotating D3-brane}

We now consider the nonextremal rotating D3-brane solution of type
IIB theory. This solution is nonsupersymmetric and is conjectured
to correspond to the Coulomb branch of a field theory at finite
temperature. For the particular case in which all three angular
momenta are equal, the gravity solution is given by \cite{larsen}
\bea ds_{10}^2 &=& \fft{Lr^2}{R^2}\, \Big[ -\fft{R^4}{L^2
r^4}\,dt^2+dx_i^2 + \fft{2m}{r^4} \Big( \ell^{-1}
\cosh\delta\,dt-B_{(1)}\Big) ^2 \Big]
+\fft{R^2 L}{r^2}\, \fft{dr^2}{L^3-2m/r^4}\nnn\\
&+& R^2\, [d\alpha^2+s_{\alpha}^2\,d\theta^2 +c_{\alpha}^2\,
(d\psi-d\varphi_2)^2 +s_{\alpha}^2 s_{\theta}^2\,
(d\psi-d\varphi_1)^2 +s_{\alpha}^2 c_{\theta}^2\,
(d\psi+d\varphi_1+d\varphi_2)^2]\,,\nnn\\
F_{(5)} &=& dC_{(4)}+\ast\,dC_{(4)}\,,\nnn\\ C_{(4)} &=& \Big(
1-\fft{L^2 r^4}{R^4}\Big) \sinh^{-1}\delta\,\, d^3x\wedge
(\cosh\delta\,dt-\ell\,B_{(1)})\,, \label{nonextremal} \eea
where
\be B_{(1)}= s_{\alpha}^2 c_{2\theta}\,d\varphi_1 +(s_{\alpha}^2
c_{\theta}^2-c_{\alpha}^2) \,d\varphi_2 +d\psi\,, \ee
and
\be L=1+\fft{\ell^2}{r^2}\,. \ee
We have defined $R^4\equiv 2m\,\sinh^2\delta$. The quantities $R$,
$\ell$ and $m$ parameterize the D3-brane charge, angular momenta
and energy above extremality, respectively. The extremal limit
$m=0$ corresponds to a static distribution of D3-branes, as was
discussed in section 2. In particular, here the D3-branes would be
uniformly distributed on the surface of a 5-sphere.

For $\ell^4<\ft{8}{27}m$, there is an event horizon around the
singularity. This has been interpreted as the high-density and
high-temperature deconfined phase of the dual field theory. For
$\ell^4>\ft{8}{27}m$, the Hawking temperature vanishes and there
is no longer an event horizon hiding the singularity. There are
indications that this marks the transition from the deconfined
high-density phase of the dual field theory to the Coulomb phase
at finite density \cite{evans}.

The metric can be written in the form given by (\ref{metric}) with
\bea F^{-1/3} g_{\mu\nu} dx^{\mu} dx^{\nu} &=& c_9\,dt^2+c_7^{-1}
(d\psi+c_8\,dt)^2 + \fft{Lr^2}{R^2}\,dx_i^2\nnn\\ &+& \fft{R^2
L}{r^2}\, \fft{dr^2}{L^3-2m/r^4} +R^2\,
[d\alpha^2+s_{\alpha}^2\,d\theta^2]\,, \eea
where
\bea F &=& \fft{1}{\sqrt{c_1 c_4}}\,,\qquad
\Delta=\fft{c_1}{c_4}\,,\nnn\\
C &=& -c_1\,R^2 s_{\alpha}^2 \Big( c_{\theta}^2
+\fft{L}{\sinh^2\delta\,r^2} c_{2\theta} (s_{\alpha}^2
c_{\theta}^2 -c_{\alpha}^2)\Big)\,, \eea
and
\bea D\varphi_1 &=& d\varphi_1+(c_2+C\,c_5)\,d\psi
+(c_3+C\,c_6)\,dt\,,\nnn\\
D\varphi_2 &=& d\varphi_2+c_5\,d\psi+c_6\,dt\,. \eea
The $c_i$ functions are given by (\ref{nonextremalci}). The
deformed solution can be expressed as (\ref{deformed}) together
with the $F_{(5)}$ given in (\ref{nonextremal}). One can
straightforwardly dualize $F_{(5)}$ to obtain the explicit
expression for $d{\cal A}_{(2)}$ which is too long to present.

\section{Further directions}

We have considered supersymmetric deformations of various type IIB
supergravity backgrounds which exhibit flows away from an
asymptotically locally $AdS_5\times S^5$ fixed point. This
includes the near-horizon limit of a static ellipsoidal
distribution of D3-branes which, after the deformations,
corresponds to a region along the Coulomb branch of ${\cal N}=1$
super Yang Mills theory. The marginal deformation of $AdS_5\times
S^5$ considered in \cite{LM} lies in the UV limit of the
deformation of this solution, and the deformed superpotential does
not change as one moves onto the Coulomb branch. However, there
still remains a short-distance naked singularity on the surface of
D3-branes in the gravity solution. This singularity can be hidden
by an event horizon if the D3-branes rotate in the transverse
space. This breaks supersymmetry and corresponds to giving the
dual field theory at finite temperature. In order for the solution
still be supersymmetric, the D3-branes must also have a specific
amount of rotation within their worldvolume as well \cite{reall1}.
We also discuss the deformations of these solutions.

Another example of type IIB flows away from $AdS_5\times S^5$ is
provided by D3-branes wrapped on a (quotiented) hyperbolic 2-cycle
\cite{MN1}. This describes a twisted field theory \cite{twisted}
which, upon deformation, exhibits a flow ``across dimensions''
from an ${\cal N}=1$ four-dimensional superconformal field theory
to a $(2,2)$ two-dimensional superconformal field theory. The
four-dimensional field theory lives on a Minkowski$_2\times H^2$
background. We also consider the deformation of a gravity dual
which corresponds to a four-dimensional field theory which lives
on $AdS_3\times S^1$ \cite{LPV}, although the dual field theory
has not yet been identified. Note that, after the deformations,
all of these solutions still exhibit conformal fixed-points at
both ends of the flow.

For all of our examples, we could have also considered dipole
deformations by involving a worldvolume direction in the $SL(3,R)$
transformation \cite{GN}. In particular, for the cases in which
the four-dimensional field theory lives on Minkowski$_2\times H^2$
or $AdS_3\times S^1$, this could involve one direction of $H^2$ or
the $S^1$ direction, respectively. Then this is a dipole
deformation of the four-dimensional UV field theory and a marginal
deformation of the two-dimensional IR field theory. RG flows which
connect these different types of deformations are rather novel.

There are a number of other type IIB solutions which are
asymptotically $AdS_5\times S^5$ and can be deformed by the
technique of \cite{LM}. Such possibilities include a subclass of
the bubbling $AdS$ solutions found in \cite{LLM} and $AdS_5$ black
holes with fermionic hair \cite{liu}. One can also apply similar
deformations to the $AdS_3\times S^3\times S^3\times S^1$
background. Possible holographic duals of this gravity background
have been discussed in \cite{gukov}. The initial two-dimensional
field theory has {\it large} ${\cal N}=4$ superconformal symmetry,
due to the $SU(2)\times SU(2)$ R-symmetry. Depending on which
directions of the gravity dual are used for the deformation, the
deformed field theory might have ${\cal N}=(2,2)$, $(0,4)$ or {\it
small} $(4,4)$ supersymmetry.

We have also presented two types of nonsupersymmetric
generalizations of the deformations of Coulomb branch flows. There
is an ${\cal N}=1$ supersymmetric flow \cite{KW,GW} which
encompasses the ${\cal N}=4$ Coulomb branch flow \cite{larsen} and
the ${\cal N}=1$ Pilch-Warner flow \cite{PW1}. The deformation is
nonsupersymmetric, since it involves the direction associated with
the $U(1)$ R-symmetry. This is clearly not the gravity dual of an
${\cal N}=4$ super Yang-Mills theory which has undergone both
marginal and relevant deformations \cite{BJL}, since the resulting
solution would need to be supersymmetric. In fact, it would be
interesting to find the gravity dual for this, which may require a
deformation technique that does not involve T-duality. Anyway, one
can also apply this deformation procedure to the ${\cal N}=2$
supersymmetric RG flow solution found in \cite{PWapril,PWjune},
although the result will also be nonsupersymmetric. There are also
eleven-dimensional flows which can be deformed analogously, such
as the ones described in \cite{CPW,AI2001,AI2002}.

For these deformed solutions which do not preserve any
supersymmetry, it is important to study their stability. Masses
should vary smoothly with the deformation parameters. Therefore,
it seems reasonable to expect the lightest modes to become
tachyonic first. Also, for small deformations, tachyons should be
absent from the spectrum.

There is also a nonsupersymmetric generalization of the Coulomb
branch flow which corresponds to the gauge theory being at finite
temperature \cite{larsen}. The initial solution is already
nonsupersymmetric, which means there are actually three directions
which can be paired up for the deformations with a chain of
dualities and shifts. This yields a solution with six deformation
parameters which reduces to the solution discussed in
\cite{Frolov,Frolov:2005ty,Frolov:2005iq} in the UV limit. There,
it was found that a three-parameter case of the latter solution
has an associated string sigma model which is integrable. We do
not know if this property of integrability survives away from the
UV limit. The initial solution suffers from an instability for
large angular momenta. Above this threshold, the D3-branes may
split apart into fragments which move radially outwards and carry
away some of the spin. This could correspond to a spurious gauge
field vev producing a negative mass scalar term which destabilizes
the moduli space of the dual gauge theory. It would be interesting
to see if these deformations simply change the threshold angular
momenta for stability, or if the entire solution is rendered
unstable.

\vspace{.7cm}

\centerline{\bf Acknowledgments}

\indent

We would like to thank Philip Argyres, Oleg Lunin, Juan Maldacena
and Carlos Nu\~{n}ez for useful discussions. In particular, we
appreciate the detailed explanations provided by Oleg Lunin with
regards to his recent paper. The work of C.A. is supported by a
grant from the Monell Foundation through the Institute for
Advanced Study, and by grant No. R01-2006-000-10965-0 from the
Basic Research Program of the Korea Science \& Engineering
Foundation. The work of J.F.V.P. is supported by DOE grant
DOE-FG02-84ER-40153.

\appendix

\renewcommand{\thesection}{\large \bf \mbox{Appendix~}\Alph{section}}
\renewcommand{\theequation}{\Alph{section}\mbox{.}\arabic{equation}}

\section{Components of $d{\cal A}_{(2)}$ in section 3}

The terms in $F_{(5)}$ which yield contributions to the two-forms
of the deformed solution can be written in terms of $d{\cal
A}_{(2)}$, whose components are given by
\bea F_{\alpha\theta\psi} &=&
\fft{6R^4}{{\td\Delta}^{5/2}}\,s_{2\theta} s_{\alpha}^3 c_{\alpha}
(1+{\td\Delta}\,e^{-2\varphi}) \Big( c_{\alpha}^2+ e^{3\varphi}
s_{\alpha}^2\Big)^{1/2}\,,\nnn\\
F_{r\theta\psi} &=& \fft92 R^3 s_{2\theta} s_{\alpha}^3
c_{\alpha}^2 \fft{e^{\varphi/2}}{{\td\Delta}^{5/4}}
(c_{\alpha}^2+e^{3\varphi}
s_{\alpha}^2)^{-1/2} \varphi'\,,\nnn\\
F_{xy\alpha} &=& \fft{R^3}{3y^2} \fft{s_{\alpha}}{c_{\alpha}}
\fft{2A_{\psi}^1+A_{\psi}^2}{k{\td\Delta}}
(c_{\alpha}^2+e^{3\varphi} s_{\alpha}^2)^{1/2}\,,\nnn\\
F_{xy\theta} &=& \fft{2R^3}{3y^2} \fft{s_{\alpha}^2}{s_{2\theta}}
\fft{e^{-2\varphi}}{k{\td\Delta}} (c_{\alpha}^2+e^{3\varphi}
s_{\alpha}^2)^{-1/2} [a(1-c_{2\theta} A_{\psi}^1-c_{\theta}^2
A_{\psi}^2)e^{2\varphi}
+(2a-1)(1-A_{\psi}^2)e^{-4\varphi}]\,,\nnn\\
F_{y\theta \psi} &=& \fft{R^4}{2y} {\td\Delta}^{-3/2} e^{-2g}
(c_{\alpha}^2+e^{3\varphi} s_{\alpha}^2)^{-1/2} s_{\alpha}^4
c_{\alpha}^2 s_{2\theta}\, [(2a+(a-3a_1)A_{\psi}^1 c_{2\theta}
+(a-3a_2)A_{\psi}^2 c_{\theta}^2)a\,e^{2\varphi}\nnn\\ &+&
(2(1-3a) +(3a_2-a)A_{\psi}^2)(2a-1)e^{-4\varphi}]\,,\nnn\\
F_{y\alpha\psi} &=& -\fft{R^4}{4} \fft{e^{-2g}}{y} s_{\alpha}^3
c_{\alpha} s_{2\theta}^2 {\td\Delta}^{-3/2}
(c_{\alpha}^2+e^{3\varphi} s_{\alpha}^2)^{1/2} e^{2\varphi}
[2(3a_1-a)A_{\psi}^1+(3a_2-a)A_{\psi}^2] \,. \eea

\section{Details of the solutions in section 5}

\subsection{Details of the Pilch-Warner flow}

The various functions in the Pilch-Warner flow of type IIB
supergravity can be expressed in terms of the scalar fields
$\chi(r)$ and $\rho(r)$ as \cite{FGPW}
\bea
X_1(r,\theta) & \equiv & c_{\theta}^2+\rho^6 s_{\theta}^2\,,\nnn\\
X_2(r,\theta) & \equiv & c_{\theta}^2\,{\rm sech}\chi\,+ \rho^6
s_{\theta}^2\, \cosh\chi\,, \label{X} \eea
the coefficient functions for the five-form field strength are
\bea w_r(r,\theta) &=& \fft{g\,e^{4A(r)}}{8\rho^4} \cosh^2\chi\,
[c_{\theta}^2 (\cosh (2\chi)-3)+\rho^6\, (2\rho^6 s_{\theta}^2
\sinh^2\chi\,+c_{2\theta}\,-3)]\,,
\nnn\\
w_{\theta}(r,\theta) &=& \fft{e^{4A(r)}}{8\rho^2} s_{2\theta}
[2\cosh^2\chi+\rho^6 (\cosh (2\chi)-3)], \label{w} \eea
and the complex two-form potential involves the coefficient
functions
\bea a_1(r,\theta) &=& \fft{2}{g^2} c_{\theta}\,\tanh\chi\,,\qquad
a_2(r,\theta) = \fft{\rho^6}{g^2 X_1} c_{\theta}^2\,s_{\theta}
\,\tanh\chi\,,\nnn\\ a_3(r,\theta) &=& -\fft{2}{g^2 X_1}
c_{\theta}^2\,s_{\theta}\, \tanh\chi\,. \label{a} \eea
The complex two-form potential can be written in terms of RR and
NS-NS two-form potentials $C_{(2)}^{{\rm RR}}$ and $B_{(2)}^{{\rm
NS}}$, whose components are given by
\bea (B_{12},C_{12}) &=&
(a_3-a_2)s_{\alpha}(s_{\phi},c_{\phi})\,,\qquad
(B_{1\alpha},C_{1\alpha})=(a_3-a_2)(c_{\phi},-s_{\phi})\,,\nnn\\
(B_{1\phi},C_{1\phi}) &=& (a_2-a_3)c_5
s_{\alpha}(s_{\phi},c_{\phi})\,,\qquad (B_{2\phi},C_{2\phi}) =
[a_3+(a_3-a_2)(c_3-c_2 c_5)]s_{\alpha}(s_{\phi},c_{\phi})\,,\nnn\\
(B_{2\alpha},C_{2\alpha}) &=& a_2
(-c_{\phi},s_{\phi})\,c_{\alpha}\,,\qquad
(B_{2\theta},C_{2\theta})=a_1 s_{\alpha} (c_{\phi},-s_{\phi})\,,
\label{BC} \\
({\td b}_{\phi\alpha},{\td c}_{\phi\alpha}) &=& -({\td
b}_{\alpha\phi},{\td c}_{\alpha\phi}) =\fft12
\left[-2a_3+(a_3-a_2)(c_3-c_2 c_5)- a_2
c_5\,c_{\alpha}\right](c_{\phi},-s_{\phi})\,, \nonu
\\
({\td b}_{\theta\alpha},{\td c}_{\theta\alpha}) &=& -({\td
b}_{\alpha\theta},{\td
c}_{\alpha\theta})=a_1\,(s_{\phi},c_{\phi})\,,\qquad ({\td
b}_{\theta\phi},{\td c}_{\theta\phi}) = -({\td
b}_{\phi\theta},{\td c}_{\phi\theta})=a_1 c_5\,s_{\alpha}
(c_{\phi},-s_{\phi})\,, \nonu \eea
where all unspecified components are zero. Notice that
$C_{(2)}^{{\rm RR}}(\phi)=B_{(2)}^{{\rm NS}}(\phi+\pi/2)$. The
$c_i$ functions are given by
\bea c_1(r,\theta) &=& \fft{\cosh\chi}{X_1^2 X_2 \rho^6} (X_1^2
\rho^6
c_{\theta}^2+f^2 s_{\theta}^2)\,,\nnn\\
c_2(r,\theta)  &=& \fft{c_{\theta}^2\,
c_{\alpha}\cosh\chi\,}{X_1^2 X_2 c_1}
(X_1^2+f\,s_{\theta}^2\sinh\chi\,\tanh\chi\,)\,,\nnn\\
c_3(r,\theta)  &=& \fft{2f\,s_{\theta}^2\cosh\chi\,}{X_1^2 \rho^6
c_1}\,,\nnn\\
c_4(r,\theta)  &=& \fft{c_{\theta}^2\,s_{\alpha}^2}{X_1}+
\fft{c_{\theta}^2\,c_{\alpha}^2\cosh\chi\,}{X_1^2 X_2}
\left(X_1^2+\fft{\rho^6}{4} s_{
2\theta}^2\sinh^2\chi\,\tanh^2\chi\,\right)-c_1 c_2^2\,,\nnn\\
c_5(r,\theta)  &=& \fft{s_{2\theta}^2\,c_{\alpha}\,\sinh^2\chi\,
}{2X_1^2
c_4} -\fft{c_1 c_2 c_3}{c_4}\,,\nnn\\
c_6(r,\theta)  &=& \fft{4X_2 s_{\theta}^2\cosh\chi\,} {X_1^2
\rho^6} -c_1 c_3^2 -c_4 c_5^2\,,
\nonu \\
f(r,\theta)  & \equiv & \rho^6 c_{\theta}^2\sinh\chi\,
\tanh\chi\,-2X_2\,. \label{ci} \eea

\subsection{Metric functions for nonextremal rotating D3-brane}

The deformed near-horizon metric for a nonextremal rotating
D3-brane can be expressed in terms of the functions
\bea c_1^{-1} &=& R^2 s_{\alpha}^2 \Big( 1+
\fft{L}{\sinh^2\delta\, r^2} s_{\alpha}^2
c_{2\theta}^2\Big)\,,\nnn\\
c_2 &=& c_1\,R^2 s_{\alpha}^2 c_{2\theta}\, \Big(
1+\fft{L}{\sinh^2\delta\,r^2}\Big)\,,\nnn\\ c_3 &=& -\fft{c_1\,R^2
 \cosh\delta}{\sinh^2\delta\,r^2}\, s_{\alpha}^2
c_{2\theta}\,,\nnn\\
c_4^{-1} &=& c_1\,R^4 s_{\alpha}^2 \Big( c_{\alpha}^2+
s_{\alpha}^2 s_{\theta}^2 c_{\theta}^2+
\fft{L}{\sinh^2\delta\,r^2} (s_{\alpha}^2
c_{\theta}^2-c_{\alpha}^2) [s_{\alpha}^2
c_{\theta}^2 (1-2c_{2\theta})-c_{\alpha}^2]\Big)\,,\nnn\\
c_5 &=& c_4\, R^2 \Big( 1+\fft{L}{\sinh^2\delta\,r^2}\Big)
(s_{\alpha}^2 c_{\theta}^2-c_{\alpha}^2+C\, s_{\alpha}^2
c_{2\theta})\,,\nnn\\
c_6 &=& \fft{c_4\, R^2 \cosh\delta}{\sinh^2\delta\, r^2}\,
(c_{\alpha}^2-s_{\alpha}^2 c_{\theta}^2+C\, s_{\alpha}^2
c_{2\theta})\,,\nnn\\
c_7^{-1} &=& R^2 \Big( 1+\fft{L}{\sinh^2\delta\,r^2}\Big)
(2-3s_{\alpha}^2 c_{\theta}^2-C\, s_{\alpha}^2
c_{2\theta})\,,\nnn\\
c_8 &=& -\Big( \fft{R^2 \cosh\delta}{\sinh^2\delta\,r^2} +c_4^{-1}
c_5 c_6+c_1^{-1} c_2 c_3\Big) c_7\,,\nnn\\
c_9 &=& \fft{R^2}{L\,\sinh^2\delta\,r^2}-c_1^{-1} c_3^2- c_4^{-1}
c_6^2-c_7^{-1} c_8^2\,. \label{nonextremalci} \eea


\begin{thebibliography}{99}

\bibitem{Malda1997}
J.~M.~Maldacena, ``The large N limit of superconformal field
theories and supergravity,'' Adv.\ Theor.\ Math.\ Phys.\  {\bf 2},
231 (1998); Int.\ J.\ Theor.\ Phys.\  {\bf 38}, 1113
(1999)\newline [arXiv:hep-th/9711200].

\bibitem{GKP}
S.~S.~Gubser, I.~R.~Klebanov and A.~M.~Polyakov, ``Gauge theory
correlators from non-critical string theory,'' Phys.\ Lett.\ B
{\bf 428}, 105 (1998) [arXiv:hep-th/9802109].

\bibitem{Witten1998}
E.~Witten, ``Anti-de Sitter space and holography,'' Adv.\ Theor.\
Math.\ Phys.\  {\bf 2}, 253 (1998) [arXiv:hep-th/9802150].

\bibitem{LM}
O.~Lunin and J.~Maldacena, ``Deforming field theories with U(1) x
U(1) global symmetry and their gravity duals,'' JHEP {\bf 0505},
033 (2005) [arXiv:hep-th/0502086].

\bibitem{LS}
R.~G.~Leigh and M.~J.~Strassler, ``Exactly marginal operators and
duality in four-dimensional N=1 supersymmetric gauge theory,''
Nucl.\ Phys.\ B {\bf 447}, 95 (1995) [arXiv:hep-th/9503121].

\bibitem{AV}
C.~Ahn and J.~F.~V\'{a}zquez-Poritz, ``Marginal deformations with
$U(1)^3$ global symmetry,'' JHEP {\bf 0507}, 032 (2005)
[arXiv:hep-th/0505168].

\bibitem{Ypq1}
J.~P.~Gauntlett, D.~Martelli, J.~Sparks and D.~Waldram,
``Supersymmetric AdS(5) solutions of M-theory,'' Class.\ Quant.\
Grav.\  {\bf 21}, 4335 (2004) [arXiv:hep-th/0402153].

\bibitem{Ypq2}
J.~P.~Gauntlett, D.~Martelli, J.~Sparks and D.~Waldram,
``Sasaki-Einstein metrics on S(2) x S(3),'' Adv. Theor. Math.
Phys. {\bf 8}, 711 (2004) [arXiv:hep-th/0403002].

\bibitem{Lpqr1}
M.~Cveti\v{c}, H.~L\"{u}, D.~N.~Page and C.~N.~Pope, ``New
Einstein-Sasaki Spaces in Five and Higher Dimensions,''
[arXiv:hep-th/0504225].

\bibitem{Lpqr2}
D.~Martelli and J.~Sparks, ``Toric Sasaki-Einstein metrics on S(2)
x S(3),'' Phys. Lett. {\bf B621}, 208 (2005)
[arXiv:hep-th/0505027].

\bibitem{klebstrass}
I.R. Klebanov and M.J. Strassler, ``Supergravity and a confining
gauge theory: Duality cascades and $\chi$SB-resolution of naked
singularities,'' JHEP {\bf 0008}, 052 (2000)
[arXiv:hep-th/0007191].

\bibitem{maldnun}
J.M. Maldacena and C. Nu\~{n}ez, ``Towards the large $N$ limit of
pure ${\cal N}=1$ super Yang Mills,'' Phys. Rev. Lett. {\bf 86},
588 (2001) [arXiv:hep-th/0008001].

\bibitem{GN}
U.~Gursoy and C.~Nunez, ``Dipole deformations of N = 1 SYM and
supergravity backgrounds with U(1) x U(1) global symmetry,''
[arXiv:hep-th/0505100].

\bibitem{GMSW03}
J.~P.~Gauntlett, D.~Martelli, J.~F.~Sparks and D.~Waldram, ``A new
infinite class of Sasaki-Einstein manifolds,''
[arXiv:hep-th/0403038].

\bibitem{CLPP}
W.~Chen, H.~L\"{u}, C.~N.~Pope and J.~F.~V\'{a}zquez-Poritz, ``A
note on Einstein-Sasaki metrics in $D >= 7$,'' Class.\ Quant.\
Grav.\ {\bf 22}, 3421 (2005) [arXiv:hep-th/0411218].

\bibitem{GLMW}
J.~P.~Gauntlett, S.~Lee, T.~Mateos and D.~Waldram, ``Marginal
deformations of field theories with AdS(4) duals,''
[arXiv:hep-th/0505207].

\bibitem{tenauthors} M. Cveti\v{c}, M.J. Duff, P. Hoxha, James T. Liu,
H. L\"u, J.X. Lu, R. Martinez-Acosta, C.N. Pope, H. Sati, Tuan A.
Tran, ``Embedding AdS black holes in ten and eleven dimensions,''
Nucl.\ Phys.\ {\bf B558} 96 (1999), [arXiv:hep-th/9903214].

\bibitem{larsen} P. Kraus, F. Larsen and S.P. Trivedi, ``The
Coulomb branch of gauge theory from rotating branes,'' JHEP {\bf
9903} 003 (1999) [arXiv:hep-th/9811120].

\bibitem{twisted}
M. Bershadsky, C. Vafa and V. Sadov, ``D-branes and topological
field theories ,'' Nucl. Phys. {\bf B463}, 420 (1996)
[arXiv:hep-th/9511222].

\bibitem{MN1}
J. Maldacena and C. Nu\~{n}ez, ``Supergravity description of field
theories on curved manifolds and a no go theorem,'' Int. J. Mod.
Phys. {\bf A16}, 822 (2001) [arXiv:hep-th/0007018].

\bibitem{reall1} J.B. Gutowski and H.S. Reall, ``Supersymmetric
AdS5 black holes,'' JHEP {\bf 0402} 006 (2004)
[arXiv:hep-th/0401042].

\bibitem{reall2} J.B. Gutowski and H.S. Reall, ``General supersymmetric
AdS5 black holes,'' JHEP {\bf 0404} 048 (2004)
[arXiv:hep-th/0401129].

\bibitem{gauntlett} J.P. Gauntlett and J.B. Gutowski, ``All
supersymmetric solutions of minimal gauged supergravity in five
dimensions,'' Phys. Rev. {\bf D68} 105009 (2003)
[arXiv:hep-th/0304064].

\bibitem{gauntlett1} J.P. Gauntlett, J.B. Gutowski and N.V.
Suryanarayana, ``A deformation of $AdS_5\times S^5$,'' Class.
Quant. Grav. {\bf 21} 5021 (2004), [arXiv:hep-th/0406188].

\bibitem{sabra}
A.H. Chamseddine and W.A. Sabra, ``Magnetic strings in
five-dimensional gauged supergravity theories,'' Phys. Lett. {\bf
B477}, 329 (2000) [arXiv:hep-th/9911195].

\bibitem{KS}
D.~Klemm and W.~A.~Sabra, ``Supersymmetry of black strings in D =
5 gauged supergravities,'' Phys.\ Rev.\ D {\bf 62}, 024003 (2000)
[arXiv:hep-th/0001131].

\bibitem{LPV}
H.~L\"{u}, C.~N.~Pope and J.~F.~V\'{a}zquez-Poritz, ``From AdS
black holes to supersymmetric flux-branes,'' Nucl.\ Phys.\ B {\bf
709}, 47 (2005) [arXiv:hep-th/0307001].

\bibitem{jones}
D. Astefanesei and G.C. Jones, ``S-branes and (anti-)bubbles in
(A)dS space,'' JHEP {\bf 0506}, 037 (2005) [arXiv:hep-th/0502162].

\bibitem{lin}
H. Lin and J. Maldacena, ``Fivebranes from gauge theory,''
[arXiv:hep-th/0509235].

\bibitem{KW}
A.~Khavaev and N.~P.~Warner, ``An N = 1 supersymmetric Coulomb
flow in IIB supergravity,'' Phys.\ Lett.\ B {\bf 522}, 181 (2001)
[arXiv:hep-th/0106032].

\bibitem{GW}
C.~N.~Gowdigere and N.~P.~Warner, ``Holographic Coulomb branch
flows with N = 1 supersymmetry,'' [arXiv:hep-th/0505019].

\bibitem{PWfeb2000}
K.~Pilch and N.~P.~Warner, ``A new supersymmetric compactification
of chiral IIB supergravity,'' Phys.\ Lett.\ B {\bf 487}, 22 (2000)
[arXiv:hep-th/0002192].

\bibitem{PW1}
K.~Pilch and N.~P.~Warner, ``N = 1 supersymmetric renormalization
group flows from IIB supergravity,'' Adv.\ Theor.\ Math.\ Phys.\
{\bf 4}, 627 (2002) [arXiv:hep-th/0006066].

\bibitem{Frolov}
S.~Frolov, ``Lax pair for strings in Lunin Maldacena
~background,'' JHEP {\bf 0505}, 069 (2005) [arXiv:hep-th/0503201].

\bibitem{Frolov:2005ty}
S.~A.~Frolov, R.~Roiban and A.~A.~Tseytlin, ``Gauge - string
duality for superconformal deformations of N = 4 super Yang-Mills
theory,'' [arXiv:hep-th/0503192].

\bibitem{Frolov:2005iq}
S.A. Frolov, R. Roiban and A.A. Tseytlin, ``Gauge string duality
for (non)supersymmetric deformations of N=4 Super Yang-Mills
theory,'' [arXiv:hep-th/0507021].

\bibitem{youm}
M. Cveti\v{c} and D. Youm, ``Rotating intersecting M-branes,''
Nucl. Phys. {\bf B499}, 253 (1997) [arXiv:hep-th/9612229].

\bibitem{gubser}
M. Cveti\v{c} and S.S. Gubser, ``Phases of R-charged black holes,
spinning branes and strongly-coupled gauge theories,'' JHEP {\bf
9904}, 024 (1999) [arXiv:hep-th/9902195].

\bibitem{behrndt}
K. Behrndt, A.H. Chamseddine and W.A. Sabra, ``BPS black holes in
$N=2$ five-dimensional AdS supergravity,'' Phys. Lett. {\bf B442},
97 (1998) [arXiv:hep-th/9807187].

\bibitem{BLjan}
D.~Berenstein and R.~G.~Leigh, ``Discrete torsion, AdS/CFT and
duality,'' JHEP {\bf 0001}, 038 (2000) [arXiv:hep-th/0001055].

\bibitem{BJL}
D.~Berenstein, V.~Jejjala and R.~G.~Leigh, ``Marginal and relevant
deformations of N = 4 field theories and non-commutative moduli
spaces of vacua,'' Nucl.\ Phys.\ B {\bf 589}, 196 (2000)
[arXiv:hep-th/0005087].

\bibitem{BJLjune}
D.~Berenstein, V.~Jejjala and R.~G.~Leigh, ``Noncommutative moduli
spaces and T duality,'' Phys.\ Lett.\ B {\bf 493}, 162 (2000)
[arXiv:hep-th/0006168].

\bibitem{BS99}
I.~Bakas and K.~Sfetsos, ``States and curves of five-dimensional
gauged supergravity,'' Nucl.\ Phys.\ B {\bf 573}, 768 (2000)
[arXiv:hep-th/9909041].

\bibitem{CLV}
S.~Cucu, H.~L\"{u} and J.~F.~V\'{a}zquez-Poritz, ``Interpolating
from AdS(D-2) x S(2) to AdS(D),'' Nucl.\ Phys.\ B {\bf 677}, 181
(2004) [arXiv:hep-th/0304022].

\bibitem{FGPW}
D.~Z.~Freedman, S.~S.~Gubser, K.~Pilch and N.~P.~Warner,
``Renormalization group flows from holography supersymmetry and a
c-theorem,'' Adv.\ Theor.\ Math.\ Phys.\  {\bf 3}, 363 (1999)
[arXiv:hep-th/9904017].

\bibitem{Pal:2005nr}
S.~Pal, ``$\beta$-deformations, potentials and KK modes,''
[arXiv:hep-th/0505257].

\bibitem{evans}
N. Evans and J. Hockings, ``$N=4$ Super Yang Mills at finite
density: the naked truth,'' JHEP {\bf 0207}, 070 (2002)
[arXiv:hep-th/0205082].

\bibitem{LLM}
H. Lin, O. Lunin and J. Maldacena, ``Bubbling AdS space and 1/2
BPS geometries,'' JHEP {\bf 0410}, 025 (2004)
[arXiv:hep-th/0409174].

\bibitem{liu}
B.A. Burrington, J.T. Liu and W.A. Sabra, ``$AdS_5$ black holes
with fermionic hair,'' Phys. Rev. {\bf D71}, 105015 (2005)
[arXiv:hep-th/0412155].

\bibitem{gukov}
S. Gukov, E. Martinec, G. Moore and A. Strominger, ``The search
for a holographic dual to $AdS_3\times S^3\times S^3\times S^1$,''
[arXiv:hep-th/0403090].

\bibitem{PWapril}
K.~Pilch and N.~P.~Warner, ``N = 2 supersymmetric RG flows and the
IIB dilaton,'' Nucl.\ Phys.\ B {\bf 594}, 209 (2001)
[arXiv:hep-th/0004063].

\bibitem{PWjune}
K.~Pilch and N.~P.~Warner, ``Generalizing the N = 2 supersymmetric
RG flow solution of IIB supergravity,'' Nucl.\ Phys.\ B {\bf 675},
99 (2003) [arXiv:hep-th/0306098].

\bibitem{CPW}
R.~Corrado, K.~Pilch and N.~P.~Warner, ``An N = 2 supersymmetric
membrane flow,'' Nucl.\ Phys.\ B {\bf 629}, 74 (2002)
[arXiv:hep-th/0107220].

\bibitem{AI2001}
C.~Ahn and T.~Itoh, ``An N = 1 supersymmetric G(2)-invariant flow
in M-theory,'' Nucl.\ Phys.\ B {\bf 627}, 45 (2002)
[arXiv:hep-th/0112010].

\bibitem{AI2002}
C.~Ahn and T.~Itoh, ``The 11-dimensional metric for AdS/CFT RG
flows with common SU(3) invariance,'' Nucl.\ Phys.\ B {\bf 646},
257 (2002) [arXiv:hep-th/0208137].

\end{thebibliography}
\end{document}